\documentclass[prd,preprintnumbers,showpacs,12pt]{revtex4}
\usepackage{epsfig}
\usepackage{graphicx}
\usepackage {axodraw}

\usepackage{bm}
\usepackage{amsmath}
\usepackage{amssymb}
\usepackage{amscd}
\usepackage{latexsym}

\newcommand{\be}{\begin{equation}}

\newcommand{\ee}{\end{equation}}
\newcommand{\bea}{\begin{eqnarray}}
\newcommand{\eea}{\end{eqnarray}}

\newcommand{\nn}{\nonumber}

\def\Black{}
 \def\AliasBlue{}
 \def\Blue{}
 \def\Brown{}

\begin{document}

\newcommand{\bra}[1]{\langle #1|}
\newcommand{\ket}[1]{|#1\rangle}
\newcommand{\braket}[2]{\langle #1|#2\rangle}
\newcommand{\tr}{\textrm{Tr}}
\newcommand{\lag}{\mathcal{L}}
\newcommand{\mbf}[1]{\mathbf{#1}}
\newcommand{\desl}{\slashed{\partial}}
\newcommand{\Desl}{\slashed{D}}
%\preprint{{\bf DFF-413/05/04}}

\renewcommand{\bottomfraction}{0.7}
\newcommand{\epsi}{\varepsilon}

\newcommand{\nl}{\nonumber \\}
\newcommand{\tc}[1]{\textcolor{#1}}
\newcommand{\sla}{\not \!}
\newcommand{\spinor}[1]{\left< #1 \right>}
\newcommand{\cspinor}[1]{\left< #1 \right>^*}
\newcommand{\Log}[1]{\log \left( #1\right) }
\newcommand{\Logq}[1]{\log^2 \left( #1\right) }
\newcommand{\mr}[1]{\mathrm{#1}}
\newcommand{\cw}{c_\mathrm{w}}
\newcommand{\sw}{s_\mathrm{w}}
\newcommand{\ct}{c_\theta}
\newcommand{\st}{s_\theta}
\newcommand{\gt}{{\tilde g}}
\newcommand{\gtp}{{{\tilde g}^\prime}}
\renewcommand{\i}{\mathrm{i}}
\renewcommand{\Re}{\mathrm{Re}}
\newcommand{\yText}[3]{\rText(#1,#2)[][l]{#3}}
\newcommand{\xText}[3]{\put(#1,#2){#3}}

%\maketitle

\def\to{\rightarrow}
\def\ptl{\partial}
\def\beq{\begin{equation}}
\def\eeq{\end{equation}}
\def\bea{\begin{eqnarray}}
\def\eea{\end{eqnarray}}
\def\nn{\nonumber}
\def\half{{1\over 2}}
\def\rhalf{{1\over \sqrt 2}}
\def\calo{{\cal O}}
\def\call{{\cal L}}
\def\calm{{\cal M}}
\def\del{\delta}
\def\eps{\epsilon}
\def\lam{\lambda}

\def\anti{\overline}
\def\delfac{\sqrt{{2(\del-1)\over 3(\del+2)}}}
\def\heff{h'}
\def\square{\boxxit{0.4pt}{\fillboxx{7pt}{7pt}}\hspace*{1pt}}
    \def\boxxit#1#2{\vbox{\hrule height #1 \hbox {\vrule width #1
             \vbox{#2}\vrule width #1 }\hrule height #1 } }
    \def\fillboxx#1#2{\hbox to #1{\vbox to #2{\vfil}\hfil}   }

\def\braket#1#2{\langle #1| #2\rangle}
\def\gev{~{\rm GeV}}
\def\gam{\gamma}
\def\sn{s_{\vec n}}
\def\sm{s_{\vec m}}
\def\mm{m_{\vec m}}
\def\mn{m_{\vec n}}
\def\mh{m_h}
\def\sumn{\sum_{\vec n>0}}
\def\summ{\sum_{\vec m>0}}
\def\vl{\vec l}
\def\vk{\vec k}
\def\ml{m_{\vl}}
\def\mk{m_{\vk}}
\def\gp{g'}
\def\gt{\tilde g}
\def\hw{{\hat W}}
\def\hz{{\hat Z}}
\def\ha{{\hat A}}

\def\yy{{\cal Y}_\mu}
\def\yyt{{\tilde{\cal Y}}_\mu}
\def\lq{\left [}
\def\rq{\right ]}
\def\dmu{\partial_\mu}
\def\dnu{\partial_\nu}
\def\dmus{\partial^\mu}
\def\dnus{\partial^\nu}
\def\gp{g'}
\def\gpt{{\tilde g'}}
\def\gs{g''}
\def\ggs{\frac{g}{\gs}}
\def\eps{{\epsilon}}
\def\tr{{\rm {tr}}}
\def\V{{\bf{V}}}
\def\W{{\bf{W}}}
\def\Wt{\tilde{ {W}}}
\def\Y{{\bf{Y}}}
\def\Yt{\tilde{ {Y}}}
\def\L{{\cal L}}
\def\s{s_\theta}
\def\st{s_{\tilde\theta}}
\def\c{c_\theta}
\def\ct{c_{\tilde\theta}}
\def\gt{\tilde g}
\def\et{\tilde e}
\def\At{\tilde A}
\def\Zt{\tilde Z}
\def\Wpt{{\tilde W}^+}
\def\Wmt{{\tilde W}^-}

\newcommand{\Apt}{{\tilde A}_1^+}
\newcommand{\Bpt}{{\tilde A}_2^+}
\newcommand{\Amt}{{\tilde A}_1^-}
\newcommand{\Bmt}{{\tilde A}_2^-}
\newcommand{\Wtp}{{\tilde W}^+}
\newcommand{\Atp}{{\tilde A}_1^+}
\newcommand{\Btp}{{\tilde A}_2^+}
\newcommand{\Atm}{{\tilde A}_1^-}
\newcommand{\Btm}{{\tilde A}_2^-}
\def\mathswitchr#1{\relax\ifmmode{\mathrm{#1}}\else$\mathrm{#1}$\fi}
\newcommand{\Pe}{\mathswitchr e}
\newcommand{\Pp}{\mathswitchr {p}}
\newcommand{\PZ}{\mathswitchr Z}
\newcommand{\PW}{\mathswitchr W}
\newcommand{\PD}{\mathswitchr D}
\newcommand{\PU}{\mathswitchr U}
\newcommand{\PQ}{\mathswitchr Q}
\newcommand{\Pd}{\mathswitchr d}
\newcommand{\Pu}{\mathswitchr u}
\newcommand{\Ps}{\mathswitchr s}
\newcommand{\Pc}{\mathswitchr c}
\newcommand{\Pt}{\mathswitchr t}
\newcommand{\rd}{{\mathrm{d}}}
\newcommand{\GW}{\Gamma_{\PW}}
\newcommand{\GZ}{\Gamma_{\PZ}}
\newcommand{\GeV}{\unskip\,\mathrm{GeV}}
\newcommand{\MeV}{\unskip\,\mathrm{MeV}}
\newcommand{\TeV}{\unskip\,\mathrm{TeV}}
\newcommand{\fba}{\unskip\,\mathrm{fb}}
\newcommand{\pba}{\unskip\,\mathrm{pb}}
\newcommand{\nba}{\unskip\,\mathrm{nb}}
\newcommand{\PT}{P_{\mathrm{T}}}
\newcommand{\PTmiss}{P_{\mathrm{T}}^{\mathrm{miss}}}
\newcommand{\CM}{\mathrm{CM}}
\newcommand{\inv}{\mathrm{inv}}
\newcommand{\sig}{\mathrm{sig}}
\newcommand{\tot}{\mathrm{tot}}
\newcommand{\backg}{\mathrm{backg}}
\newcommand{\evt}{\mathrm{evt}}
% particle masses
\def\mathswitch#1{\relax\ifmmode#1\else$#1$\fi}
\newcommand{\M}{\mathswitch {M}}
\newcommand{\R}{\mathswitch {R}}
\newcommand{\TEV}{\mathswitch {TEV}}
\newcommand{\LHC}{\mathswitch {LHC}}
\newcommand{\MW}{\mathswitch {M_\PW}}
\newcommand{\MZ}{\mathswitch {M_\PZ}}
\newcommand{\Mt}{\mathswitch {M_\Pt}}
\def\si{\sigma}
\def\beqar{\begin{eqnarray}}
\def\eeqar{\end{eqnarray}}
\def\refeq#1{\mbox{(\ref{#1})}}
\def\reffi#1{\mbox{Fig.~\ref{#1}}}
\def\reffis#1{\mbox{Figs.~\ref{#1}}}
\def\refta#1{\mbox{Table~\ref{#1}}}
\def\reftas#1{\mbox{Tables~\ref{#1}}}
\def\refse#1{\mbox{Sect.~\ref{#1}}}
\def\refses#1{\mbox{Sects.~\ref{#1}}}
\def\refapps#1{\mbox{Apps.~\ref{#1}}}
\def\refapp#1{\mbox{App.~\ref{#1}}}
\def\citere#1{\mbox{Ref.~\cite{#1}}}
\def\citeres#1{\mbox{Refs.~\cite{#1}}}

\def\Black{}
 \def\AliasBlue{}
 \def\Blue{}
 \def\Brown{}

\title{$W^\prime$ production at the LHC in the 4-site Higgsless model}

\date{\today}% It is always \today, today,
\author{Elena Accomando}%
 \email{E.Accomando@soton.ac.uk}
\affiliation{NExT Institute and School of Physics and Astronomy, University of
Southampton, Highfield,
Southampton SO17 1BJ, UK}%
\author{Diego Becciolini}%
 \email{diego.becciolini@soton.ac.uk}
 \affiliation{NExT Institute and School of Physics and Astronomy, University of
Southampton, Highfield,
Southampton SO17 1BJ, UK}%
\author{Stefania De Curtis}%
 \email{decurtis@fi.infn.it}
\affiliation{Istituto Nazionale di Fisica Nucleare, Sezione di Firenze -Italy}%
\author{Daniele Dominici}%
 \email{dominici@fi.infn.it}
\affiliation{Universit\`a degli Studi di Firenze, Dip. di
Fisica e Astronomia, Firenze - Italy\\
and Istituto Nazionale di Fisica Nucleare, Sezione di Firenze - Italy}%
 \author{Luca Fedeli}%
 \email{fedeli@fi.infn.it}
\affiliation{Universit\`a degli Studi di Firenze, Dip. di
Fisica e Astronomia, Firenze - Italy\\
and Istituto Nazionale di Fisica Nucleare, Sezione di Firenze - Italy}%

\begin{abstract}
We study the phenomenology of the 4-site Higgsless model, based on the 
$SU(2)_L\times SU(2)_1\times SU(2)_2\times U(1)_Y$ gauge symmetry, at present 
colliders. The model predicts the existence of two neutral and four charged
extra gauge bosons, $Z_{1,2}$ and $W^\pm_{1,2}$. In this paper, we focus on 
the charged gauge sector. We first derive limits on $W_{1,2}$-boson masses and 
couplings to SM fermions from direct searches at the Tevatron. We then 
estimate at the 7 TeV LHC the exclusion limits with the actual L=1 fb$^{-1}$ 
and the discovery potential with the expected L=10 fb$^{-1}$. 
In contrast to the minimal (or 3-site) Higgsless model which 
predicts almost fermiophobic extra gauge bosons, the next-to-minimal (or 
4-site) Higgsless model recovers sizeable $W_{1,2}$-boson couplings to 
ordinary matter, expressing the non-fermiophobic multiresonance inner 
nature of extra-dimensional theories. 
Owing to this feature, we find that in one year from now the new heavy gauge 
bosons, $W_{1,2}^\pm$, could be discovered in the final state with an electron 
and large missing transverse energy at the 7 TeV LHC for $W_{1,2}$-boson 
masses in the TeV region, depending on model parameters.
\end{abstract}

\maketitle

\section{Introduction}

Extra gauge bosons $W^\prime$ are predicted in many models that extend the 
gauge structure of the Standard Model (SM).
Left-Right symmetric models~\cite{Pati:1974yy,Mohapatra:1975jm,Senjanovic:1975rk}, based on the enlarged symmetry $SU(2)_L\otimes SU(2)_R\otimes U(1)$, are 
an old and popular example; the discovery reach and the study of $W^\prime$ 
properties at the LHC has been recently re-investigated 
\cite{Gopalakrishna:2010xm,Frank:2010cj,Maiezza:2010ic} and bounds from early 
LHC data have been derived \cite{Nemevsek:2011hz}.
Heavy charged gauge bosons are also a natural consequence of Minimal Walking 
Technicolor (MWT) models. For a complete review see 
Refs.\cite{Andersen:2011yj,Sannino:2009za} and references therein. The LHC 
potential of detecting such particles have been extensively analysed 
\cite{Andersen:2011nk,Belyaev:2008yj}.   
 
During the last ten years, the idea of extra dimensions 
\cite{ArkaniHamed:1998rs,Antoniadis:1998ig,Randall:1999ee}
has been very fruitful in the proposal of extensions of the SM with or 
without the Higgs \cite{Csaki:2003dt,Agashe:2003zs,Csaki:2003zu,Barbieri:2003pr,Nomura:2003du,Cacciapaglia:2004jz,Cacciapaglia:2004rb,Contino:2006nn}. 
In these models, the $W^\prime$ bosons emerge as Kaluza Klein excitations of 
the SM $W$-boson. The possibility to discover them at the LHC has been 
investigated many years ago \cite{Accomando:1999sj}. Recently, a more refined 
analysis has been performed \cite{Boos:2011ib}. New peculiar LHC signals from 
extra charged and neutral gauge bosons have been also studied within the RS1 
models with gauge bosons in the five-dimensional 
bulk and fermions on the UV brane (except for the third generation quark) 
\cite{Agashe:2007ki,Agashe:2008jb}.
In these models EWPT (Electroweak Precision Tests) favor masses of the order 
of 2-3 TeV and therefore integrated luminosity of order 100 fb$^{-1}$ are 
required to observe the lightest states in final state top-bottom quark pairs.

The five-dimensional models  can also be deconstructed to the usual 
four-dimensional space-time
\cite{ArkaniHamed:2001ca,Arkani-Hamed:2001nc,Hill:2000mu,Cheng:2001vd,Abe:2002rj,Falkowski:2002cm,Randall:2002qr,Son:2003et,deBlas:2006fz}, where they are
described by a class of chiral Lagrangians with extended gauge symmetries.
Within the simplest deconstructed models (e.g. the 3-site model), the 
consistency of the $\epsilon_3(S)$ parameter with its experimental value is 
satisfied at tree level via the ideal localization of fermions 
\cite{Casalbuoni:2005rs,Chivukula:2005ji,SekharChivukula:2005cc,Bechi:2006sj,Casalbuoni:2007xn}. This makes the additional gauge bosons almost fermiophobic. 
As a consequence, the literature is mainly focused on complicated production 
channels: vector boson associated production and vector boson fusion. All 
these processes require high luminosity and high energy to be detected 
\cite{Birkedal:2004au,Belyaev:2007ss,He:2007ge}.
The inclusion of one-loop corrections to the $\epsilon_3(S)$ parameter allows 
for non vanishing couplings of the new gauge bosons to SM fermions 
\cite{Abe:2008hb}. In this case, the Drell-Yan (DY) production channel opens 
up. And, if followed by the most promising $W^\prime$ decay into $WZ$-pairs, 
it could allow to detect the new heavy gauge bosons at the 14 TeV LHC with 
less than 10 fb$^{-1}$ \cite{Ohl:2008ri,Abe:2011qe}.

The 4-site extension is instead much less bounded. The relation between 
the couplings of the two new gauge boson triplets with SM fermions is indeed 
strongly constrained by the $\epsilon_3$-parameter, but their magnitude is 
weakly limited by $\epsilon_1$ \cite{Accomando:2008jh,Accomando:2011vt}. The 
phenomenological consequence is that, while the minimal 3-site Higgsless 
model can be explored only in complex multi-particle processes (the easiest 
one being $pp\rightarrow VV\rightarrow 4f$ with $V=W,Z$), the 4-site 
Higgsless model can be tested in the more promising Drell-Yan channel with 
lepton pairs in the final state \cite{Accomando:2008jh,Accomando:2010ir}. 

This paper is devoted to a detailed study of signal and background for the 
leptonic Drell-Yan production of the two new charged gauge bosons predicted 
by the 4-site Higgsless model \cite{Accomando:2008jh,Accomando:2008dm,Accomando:2010fz,Accomando:2010ir,Cata:2009iy}. A similar analysis was recently 
performed within the Minimal Walking Technicolor 
\cite{Belyaev:2008yj,Andersen:2011nk}. Compared to the latter, the 4-site 
model differs by the nature of the two extra gauge bosons: the lighter is at 
leading order a vector particle while the heavier is an axial particle. The 
mass splitting $M_{W2}-M_{W1}$ is always positive and, oppositely to the 
Minimal Walking Technicolor, no mass spectrum inversion is possible. 
In the following, we consider final states with one isolated electron and 
large missing transverse momentum. We plan to investigate the 
$pp\to W^\prime\to WZ\to 4f$ mode in the future. In some cases, the latter 
channel is the most favorable 
for the observation of a heavy vector boson. Clearly, this is true when the 
bosonic decay modes are larger and consequently the already small branching 
ratios (BR) for the leptonic modes are further suppressed. In some models the 
leptonic modes could even be forbidden or very strongly suppressed as already 
noticed \cite{Matsuzaki:2006wn,Chivukula:2005bn,SekharChivukula:2006cg,SekharChivukula:2007ic}. 
%In the extended gauge models, while the couplings to fermions could be in 
%principle not  much different from the standard ones, the $W'WZ$ couplings 
%are in general substantially suppressed. In these cases the trilinear 
%couplings are non vanishing only because of mixing, after symmetry breaking 
%\cite{Matsuzaki:2006wn,Chivukula:2005bn,SekharChivukula:2006cg,
%SekharChivukula:2007ic,Accomando:2008jh,Accomando:2008dm,Accomando:2010fz,
%Accomando:2010ir}. The mixing are small and one can typically expect them to
%be of order $M_{W}^2/M_{W'}^2$
%\cite{Durkin:1985ev,London:1986dk,Rosner:1985dx,delAguila:1986iw,
%Franzini:1986zm,Barger:1986hd,Amaldi:1987fu,Jenkins:1987ue},
%so that the resulting BRs in $WZ$ are small. 

The $W^\prime$ boson has been recently searched at the Tevatron and the LHC 
in different final states. We make use of the leptonic DY channel analysis 
\cite{Aaltonen:2010jj,Abazov:2010ti,Chatrchyan:2011wq,Chatrchyan:2011dx}
to extract limits on the parameter space of the 4-site model and explore 
the discovery reach in the near future.

In Sections \ref{4-site} and \ref{properties}, we review the 4-site model and 
the properties of the two new charged gauge bosons. Section \ref{dy} is 
devoted to investigate the DY production at the LHC and the Tevatron in the 
leptonic channel. In Section \ref{disco}, we present exclusion and discovery 
reach and finally in Sect.~\ref{conclusions} we give our conclusions.

\section{The 4-site Higgsless model}
\label{4-site}

The 4-site Higgsless model represents the next-to-minimal extension of the 
3-site Higgsless model \cite{Chivukula:2006cg} that corresponds to a particular choice of the
BESS model \cite{Casalbuoni:1985kq,Casalbuoni:1986vq}. They both belong to the 
class of deconstructed Higgsless theories \cite{ArkaniHamed:2001ca,Arkani-Hamed:2001nc,Hill:2000mu,Cheng:2001vd,
Abe:2002rj,Falkowski:2002cm,Randall:2002qr,Son:2003et,deBlas:2006fz}. In their general 
formulation,
 these theories are based on the 
$SU(2)_L\otimes SU(2)^K\otimes U(1)_Y$ gauge symmetry, and contain K+1 
non-linear $\sigma$-model scalar fields interacting with the gauge fields, 
which trigger the spontaneous electroweak symmetry breaking. They  
constitute a viable alternative to the standard EWSB mechanism based on the 
existence of a light fundamental Higgs boson. The case K=1 corresponds to the 
minimal Higgsless model, more commonly called 3-site model. 

The 4-site Higgsless model is defined by taking K=2, and requiring the 
Left-Right (LR) symmetry in the gauge sector. More explicitly, it is a linear 
moose based on the electroweak gauge symmetry 
$SU(2)_L\otimes SU(2)_1\otimes SU(2)_2\otimes U(1)_Y$. Its theoretical 
foundations are presented in \cite{Casalbuoni:2005rs}, while some of its phenomenological 
consequences are analyzed in 
\cite{Accomando:2008jh,Accomando:2008dm,Accomando:2010ir,Accomando:2011vt}.

In the unitary gauge, the 4-site model predicts two new triplets of gauge 
bosons, which acquire mass through the same non-linear symmetry breaking 
mechanism giving mass to the SM gauge bosons. Let us denote with 
$W_{i\mu}^\pm$ and $Z_{i\mu}$ (i = 1, 2) the four charged and two neutral 
heavy resonances appearing as a consequence of the gauge group extension, and 
with $W^\pm_\mu$, $Z_\mu$ and $A_\mu$ the SM gauge bosons. Owing to its gauge 
structure, the 4-site Higgsless 
model a priory contains seven free parameters: the $SU(2)_L\otimes U(1)_Y$ 
gauge couplings, $\tilde g$ and $\tilde g'$, the extra $SU(2)_{1,2}$ gauge 
couplings that, for simplicity, we assume to be equal, $g_2= g_1$, the bare 
masses of lighter ($W_1^\pm, Z_1$) and heavier ($W_2^\pm, Z_2$) gauge boson 
triplets, $M_{1,2}$, and their bare direct couplings to SM fermions, $b_{1,2}$. 
However, their number can be reduced to four, by fixing the gauge couplings 
$\tilde g,\tilde g', g_1$ in terms of the three SM input parameters 
$e, G_F, M_Z$ which denote electric charge, Fermi constant and Z-boson mass, 
respectively. As a result, the parameter space is completely defined by  
four independent free parameters which we choose to be: $M_1$, $z$, $b_1$ and 
$b_2$, where $z=M_1/M_2$ is the ratio between the bare masses. In terms of 
these four parameters, physical masses and couplings of the extra gauge bosons 
to ordinary matter can be obtained via a complete numerical algorithm. This is 
one of the main results of \cite{Accomando:2011vt} where we have describe in full detail this
computation, which goes beyond the approximations commonly adopted in the 
literature. The outcome is the ability to reliably and accurately describe the 
full parameter space of the 4-site Higgsless model even in regions of low mass 
and high $z$ where previously used approximations would fail. In the 
following, we choose to describe the full parameter space via the physical 
observables: $M_{W1}, z, a_{W1}, a_{W2}$ which denote the mass of the lighter 
extra charged gauge boson, the ratio between bare masses (which, as shown in 
\cite{Accomando:2011vt} is a good approximation of the ratio between physical 
masses $M_{W1}/M_{W2}$), and the couplings of lighter and heavier extra 
charged gauge bosons to ordinary matter, respectively. 
 
In terms of the above quantities, the Lagrangian describing the 
interaction between gauge bosons and fermions has the following expression: 
\bea
\label{aw}
{\mathcal L}_{NC}&=&\bar{\psi} \gamma^\mu \left [- e \mathbf{Q}^f A_\mu
+ {a}_{Z}^f Z_\mu+{a}_{Z_1}^fZ_{1\mu}+ {a}_{Z_2}^fZ_{2\mu}  \right ]\psi\nn\\
{\mathcal L}_{CC}&=&\bar{\psi} \gamma^\mu T^-  \psi \left(
a_WW_{\mu}^+ +a_{W_1} W_{1\mu }^+ +a_{W_2} W_{2\mu }^+\right) + h.c.
\eea
for the neutral (NC) and charged (CC) gauge sector, respectively. In the above 
formulas, $\psi$ denotes SM quarks and leptons. These expressions will be
used later when discussing production and decay of the two extra charged gauge 
bosons in the Drell-Yan channel. 

Before performing any meaningful analysis, it is mandatory to evaluate the 
impact of the Electroweak Precision Tests on the 4-site model. In the next 
section, we therefore review the constraints on the 4-site parameter space 
coming from EWPT.
 
\subsection{EWPT bounds}
\label{sec:EWPT}

Universal electroweak radiative corrections to the precision observables 
measured by LEP, SLD and TEVATRON experiments can be efficiently quantified in 
terms of three parameters: $\epsilon_1, \epsilon_2$, and $\epsilon_3$ 
(or S, T, and U)
\cite{Peskin:1990zt,Peskin:1992sw,Altarelli:1991zd,Altarelli:1998et}. 
Besides these SM contributions, the $\epsilon_i$ (i=1,2,3) parameters allow to 
describe the low-energy effects of potential heavy new physics. For that 
reason, they are a powerful method to constrain theories beyond the SM. 
In a recent paper 
\cite{Accomando:2011vt}, we used this parametrization to extract bounds on the 
3-site and 4-site Higgsless models. 
In order to derive realistic and reliable constraints, we performed a complete 
numerical calculation of all $\epsilon_i$ (i=1,2,3) parameters at tree level, 
going beyond popular approximations used in the past, and carried out a 
combined fit to the experimental results taking into account their full 
correlation. The outcome represents a drastic update of past analysis. In the 
literature, in fact, these tree level new physics effects had been evaluated 
via an analytical truncated multiple expansion in the extra gauge coupling, 
$e/g_1$, and the direct couplings of the extra gauge bosons with SM fermions 
(or delocalization parameters), that is $b_{1,2}$ in our notation. The exact 
result we presented in \cite{Accomando:2011vt} allows one to span the full 
parameter space of the model, reliably computing also regions characterized by 
small $g_1$ (or $M_1$ ) values, and sizeable $b_{1,2}$ bare couplings where the 
common approximated expansion would fail. 

\begin{figure}[t]
\begin{center}
\unitlength1.0cm
\begin{picture}(7,4)
\put(-5.6,-4){\epsfig{file=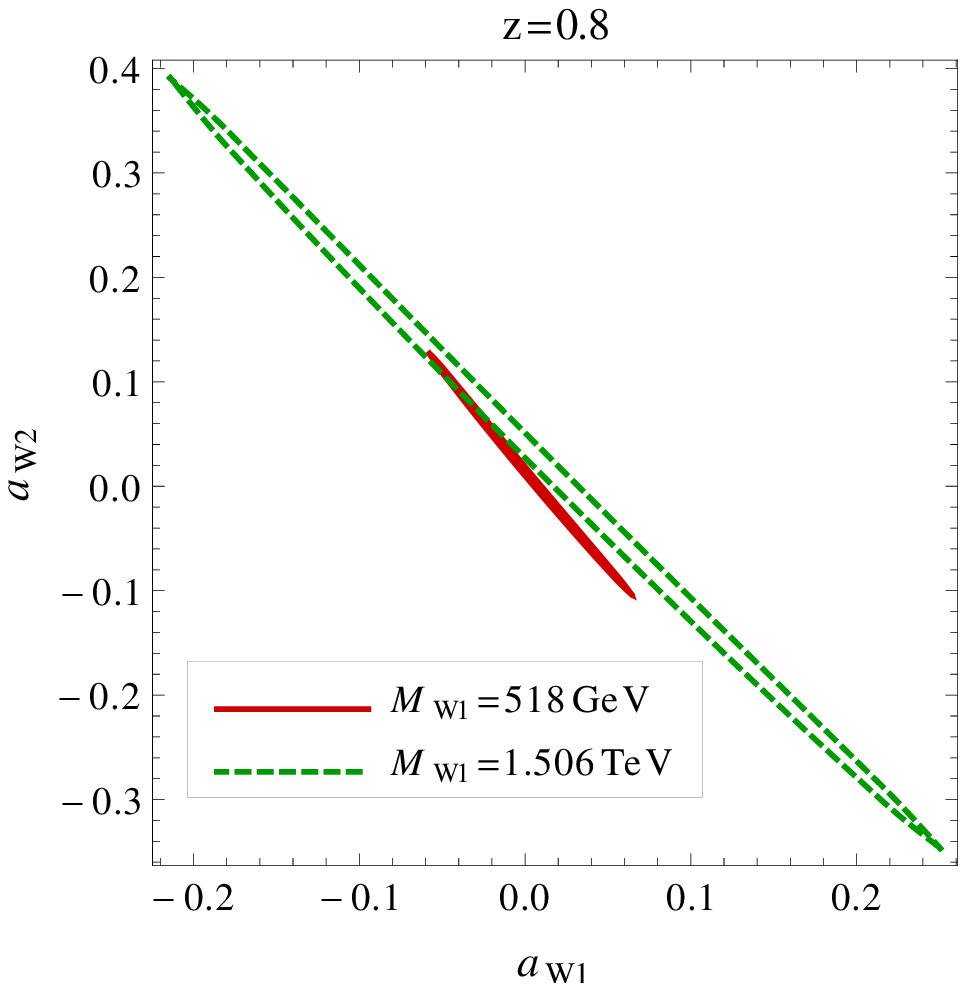,width=7.5cm}}
\put(3.5,-4){\epsfig{file=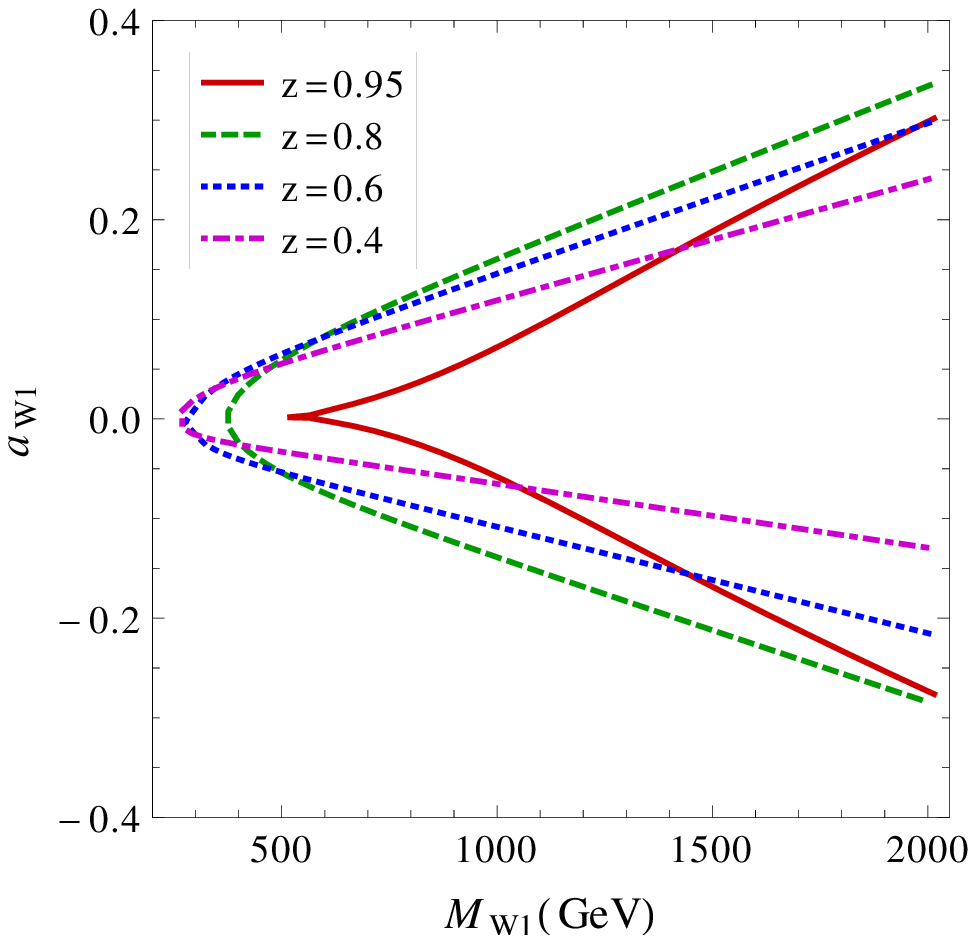,width=7.5cm}}
\end{picture}
\end{center}
\vskip 3.5cm
\caption{Left: 95$\%$ CL limits from EWPT in the $a_{W1}, a_{W2}$ plane for 
$z$=0.8 and two reference values of the lighter extra gauge boson 
mass, $M_{W1}$. Right: 95$\%$ CL EWPT bound in the parameter space given in 
terms of physical mass, $M_{W1}$ , and physical coupling between lighter 
extra charged gauge bosons and SM fermions, $a_{W1}$. 
We consider four reference $z$-values: $z$= 0.4, 0.6, 0.8 and 0.95. The 
allowed regions are delimited by the curves.}
\label{fig:EWPT_a1-a2}
\end{figure}

The major consequence of the analysis in \cite{Accomando:2011vt} is that 
both $\epsilon_1$ and $\epsilon_3$ play a fundamental role in constraining the
4-site Higgsless model ($\epsilon_1$ being usually considered sub-dominant).
While $\epsilon_3$ generates a strong correlation between the couplings of
lighter and heavier extra charged (or neutral) gauge bosons to SM fermions,
$a_{W1,W2}$, the $\epsilon_1$ parameter limits indeed their magnitude. This 
effect is shown in the left panel of  
Fig.~\ref{fig:EWPT_a1-a2}, where we plot the 95$\%$ CL limits from EWPT in the 
$a_{W1}, a_{W2}$ plane for $z$=0.8 and two $M_{W1}$ reference values.
Let us notice that the signs of the physical
fermion-boson couplings are completely arbitrary and physically irrelevant,
what only matters is their size. We introduce signs to distinguish different 
points of the parameter space \cite{Accomando:2011vt}. 

Owing to the above mentioned correlation, the number of free parameters can 
be further reduced to three. With this mild approximation, we can choose to 
describe the parameter space of the 4-site model in terms of the following 
set of physical quantities: $M_{W1}$, $a_{W1}$ and $z$. The right panel of 
Fig.~\ref{fig:EWPT_a1-a2} displays the 95$\%$ CL bounds from EWPT in the 
$M_{W1}, a_{W1}$ plane for four reference values of the $z$ parameters: 
$z$=0.4, 0.6, 0.8 and 0.95. From Fig.~\ref{fig:EWPT_a1-a2} we deduce 
that, even if constrained, the $a_{W1}$ coupling can be of the same order of 
magnitude than the corresponding SM coupling. This result is common to all
other couplings between extra gauge bosons and ordinary matter, which can be
uniquely derived from $a_{W1}$ via our complete numerical algorithm.
This is an important property which makes a very clear distinction between 4-site and 3-site model. The latter predicts indeed
 a unique gauge boson triplet, constrained to be (almost) fermiophobic in order to reconcile unitarity and EWPT bounds. This 
feature can be extrapolated from Fig.~\ref{fig:EWPT_a1-a2} by looking at the strong $z$-parameter dependence of the contours,
 which shrink with
decreasing $z$, going towards the $z\rightarrow 0$ limit where one recovers the 3-site Higgsless model. Hence, oppositely to 
the minimal model, the next-to-minimal extension (or 4-site) displays the inner extra-dimensional nature of Higgsless theories,
 which are characterized by a tower of non-fermiophobic Kaluza-Klein resonances. The 4-site model has thus the potential of 
being detected during the early stage of the LHC experiment in the Drell-Yan channel.
 
An additional information, one can extract from the right panel of 
Fig.~\ref{fig:EWPT_a1-a2}, concerns the minimum mass of the extra gauge bosons allowed by EWPT. As one can see, its value 
depends on the $z$ free parameter and can range between 250 GeV and 600 GeV 
(see Ref.\cite{Accomando:2011vt} for computational details). In this analysis 
we have not included the bounds on the trilinear gauge boson vertex, $WWZ$ 
coming from the LEP2 experiment \cite{Alcaraz:2006mx}.
The maximum allowed value for the  mass of the extra gauge bosons is instead fixed by 
the requirement of perturbative unitarity. As well known, one of the main motivations for  
Higgsless theories predicting an extended gauge sector, compared to the SM with no light elementary Higgs, is the ability to 
delay the perturbative unitarity violation up to
energy scales of the order of a few TeV. Beyond that scale, new physics should come out.
Higgsless theories must be indeed interpreted as effective low energy theories. In 
\cite{Accomando:2008jh,Accomando:2008dm,Accomando:2010ir}, all vector boson scattering (VBS) amplitudes which are
 the best smoking gun for unitarity violations are computed, with the conclusion that the 4-site Higgsless model should preserve unitarity up 
to around 3 TeV. In the following, we assume this mass validity range.

\section{Extra $W_{1,2}^\pm$-bosons: mass spectrum, decay widths and branching ratios}
\label{properties}

In this section, we summarize the main properties of the heavy charged gauge 
bosons, $W_{1,2}^\pm$, predicted by the 4-site Higgsless model. A first 
peculiarity of the 4-site model is related to the nature of the four charged 
extra gauge bosons and their mass hierarchy. The two lighter particles, 
$W_1^\pm$, are vector bosons while the heavier ones, $W_2^\pm$, are 
axial-vectors (in a 5D sense, and neglecting electroweak corrections). 
Oppositely to closely 
related models, like the walking technicolor \cite{Belyaev:2008yj}, no mass 
spectrum inversion is possible. The mass splitting, $\Delta M=M_{W2}-M_{W1}$, 
is always positive and its size depends on the free $z$-parameter: 
\be
\Delta M\sim\frac{1-z}{z}M_{W1}\qquad 0<z<1.
\label{eq:DeltaM}
\ee  
The above Eq.~(\ref{eq:DeltaM}) contains also informations on the kind of multi-resonance spectrum we might
expect. Owing to the $z$-parameter dependence, there is no fixed relation between the two charged 
gauge boson masses. We can thus have scenarios where the two charged resonances, $W_1^\pm$ and 
$W_2^\pm$, lie quite apart from each other, and portions of the parameter space in which they are (almost) degenerate. In the 
latter case, the multi-resonance distinctive signature would collapse 
into the more general single ${W^\prime}^\pm$ signal. The 4-site model would 
thus manifest a degeneracy with well known theories predicting only one 
additional pair of charged gauge bosons. The mass spectrum has both a lower 
and an upper bound, as discussed in Sect.\ref{sec:EWPT}. It lies roughly in 
the range 250 $\le M\mbox{(GeV)} \le$ 3000.

\begin{figure}[t]
\begin{center}
\unitlength1.0cm
\begin{picture}(7,4)
\put(-5.6,-4){\epsfig{file=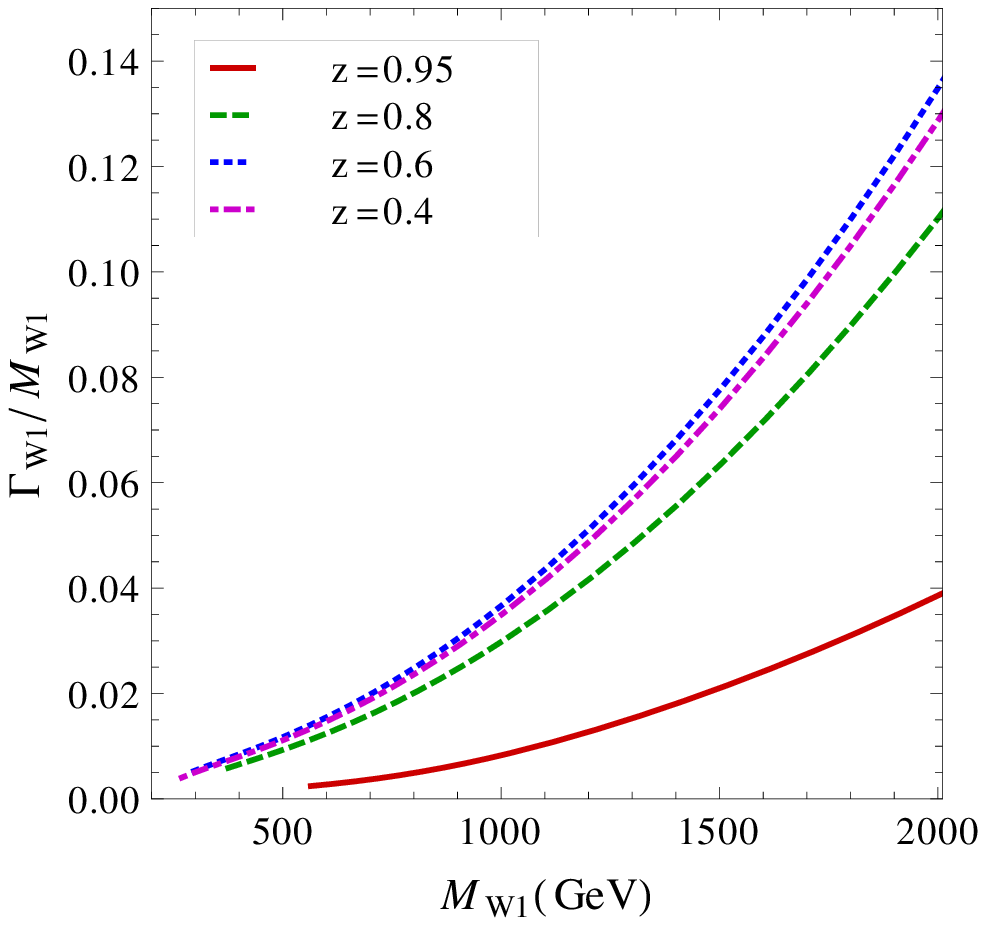,width=7.5cm}}
\put(3.5,-4){\epsfig{file=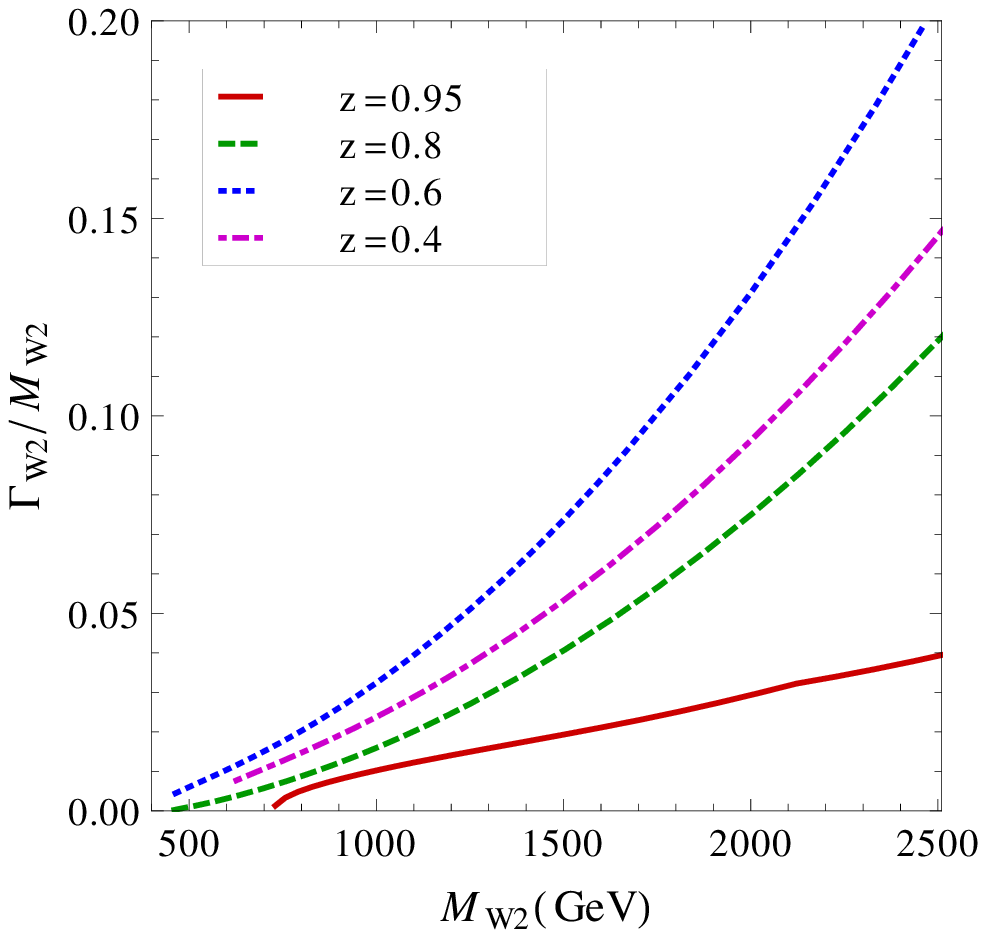,width=7.5cm}}
\end{picture}
\end{center}
\vskip 3.cm
\caption{Left: Total decay width of the lighter extra charged gauge boson, $W_1^\pm$, divided by its mass as a function of 
$M_{W1}$ for four reference values of the $z$-parameter: $z$=0.4,0.6,0.8,0.95.
Right: same for the heavier extra charged gauge boson, $W_2^\pm$.}
\label{fig:gamma12}
\end{figure}

The total widths of the heavy charged gauge bosons divided by the corresponding mass, 
$\Gamma_{W1,W2}/M_{W1,W2}$, are displayed in Fig.~\ref{fig:gamma12} as a function of the mass for four representative values
 of the $z$-parameter: $z$=0.4, 0.6, 0.8, 0.95. They have been computed by taking as $W_{1,2}^\pm ff'$ couplings those 
corresponding to the maximum value of $a_{W1}$ allowed by EWPT at fixed $z$-parameter (see contour plot in the right panel 
of Fig.\ref{fig:EWPT_a1-a2}). This maximizes the fermionic contribution to the total decay width, and it will be used later 
to show the maximal branching ratio one might expect for the $W_{1,2}^\pm$-boson decay into electrons. From 
Fig.~\ref{fig:gamma12}, one can see that both $W_{1,2}^\pm$ are very narrow for low mass values. In the low edge of the 
spectrum, the magnitude of the $W_1^\pm$-boson total width is around a few GeV as shown in the
left panel. For the $W_2^\pm$-boson in the right panel, it can go even down to a few MeV. The width is dominated by
 the $W_{1,2}^\pm$-boson decay into gauge boson pairs (when kinematically allowed) whose behaviors is proportional 
to $M_{W1,W2}^3$, it then increases with the mass up to hundreds of GeV. Since the $W_{1,2}^\pm$-boson total width can 
range between a few and hundreds of GeV, a natural question is whether one can still apply the ordinary experimental analysis
 based on the
assumption of narrow resonances. In the 4-site model indeed, both ratios $\Gamma_{W1,W2}/M_{W1,W2}$ can be bigger than the 
SM one, $\Gamma_W/M_W\simeq 2.6\%$, already for relatively low masses 
$M_{W1}\simeq$ 800 GeV, and they can approach higher values up to about 
$\Gamma_{W1,W2}/M_{W1,W2}\simeq 20\%$ at mass scales within the LHC reach, as 
shown in Fig.~\ref{fig:gamma12}. The prediction of broad resonances is a 
distinctive feature of Higgsless models, and more generally extra-dimensional
 theories \cite{Kelley:2010ap}. The benchmark model
adopted for experimental analysis, described in \cite{Altarelli:1989ff}, predicts instead a 
SM-like ${W'}^\pm$-boson purely decaying into fermion pairs. As a consequence, its width is generally expected to be very 
narrow.
In this case, signal events are rather clustered towards the Jacobian peak in the transverse mass distribution, so that 
performing a simple counting experiment in that region often gives a signal-to-background ratio that is high enough for 
detection. Broad resonances are definitely more challenging to be analyzed than narrow ones. Signal events are indeed spread
 out over a wider area, not allowing a
clear identification based on the shape. The question then becomes whether or not one can distinguish these events from the 
background or other non-resonant new physics such as contact interactions or any unshaped enhancement over the SM background. 

\begin{figure}[t]
\begin{center}
\unitlength1.0cm
\begin{picture}(7,10)
\put(-5.6,2.7){\epsfig{file=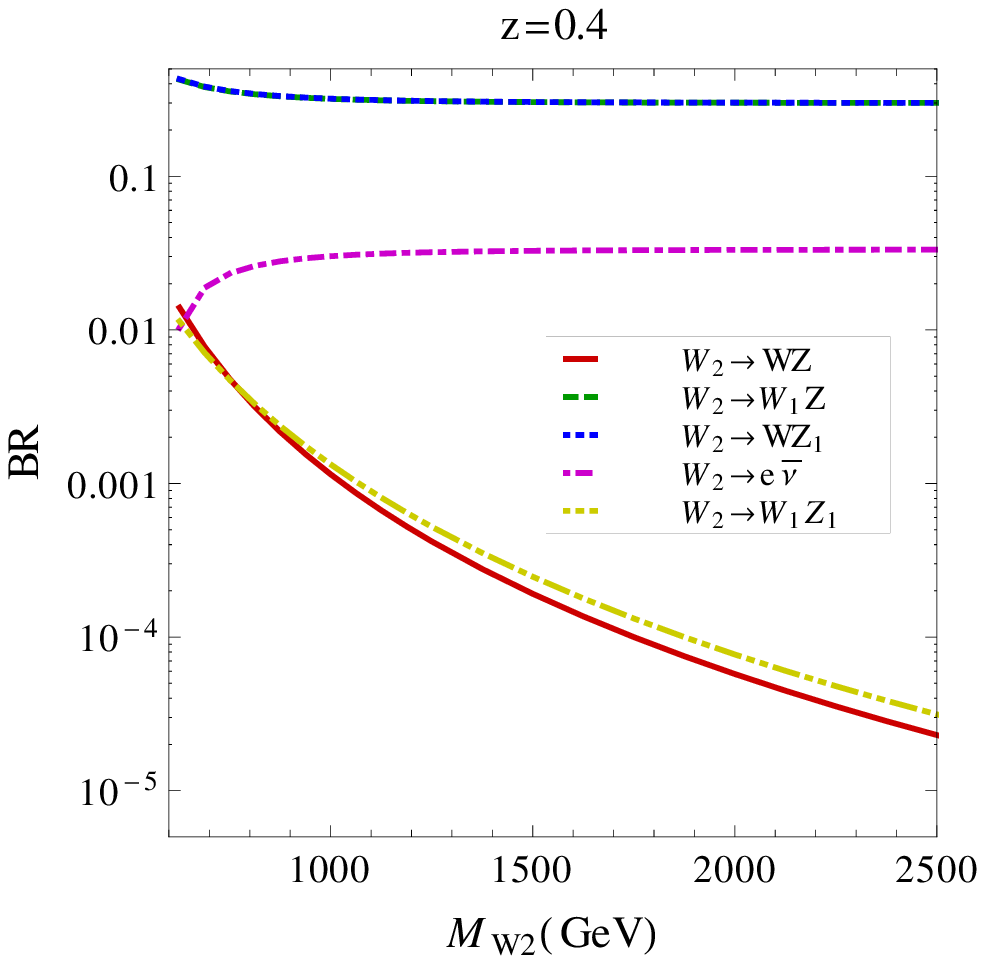,width=7.5cm}}
\put(3.5,2.7){\epsfig{file=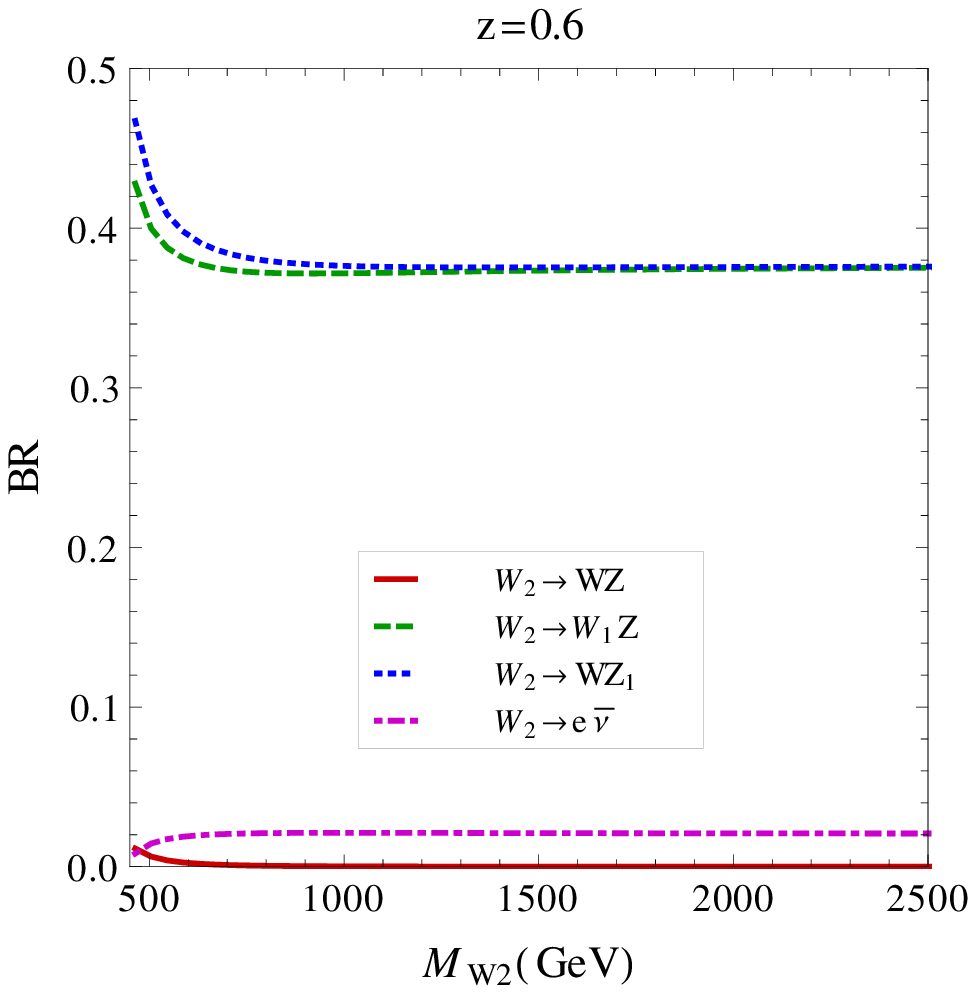,width=7.5cm}}
\put(-5.6,-4.8){\epsfig{file=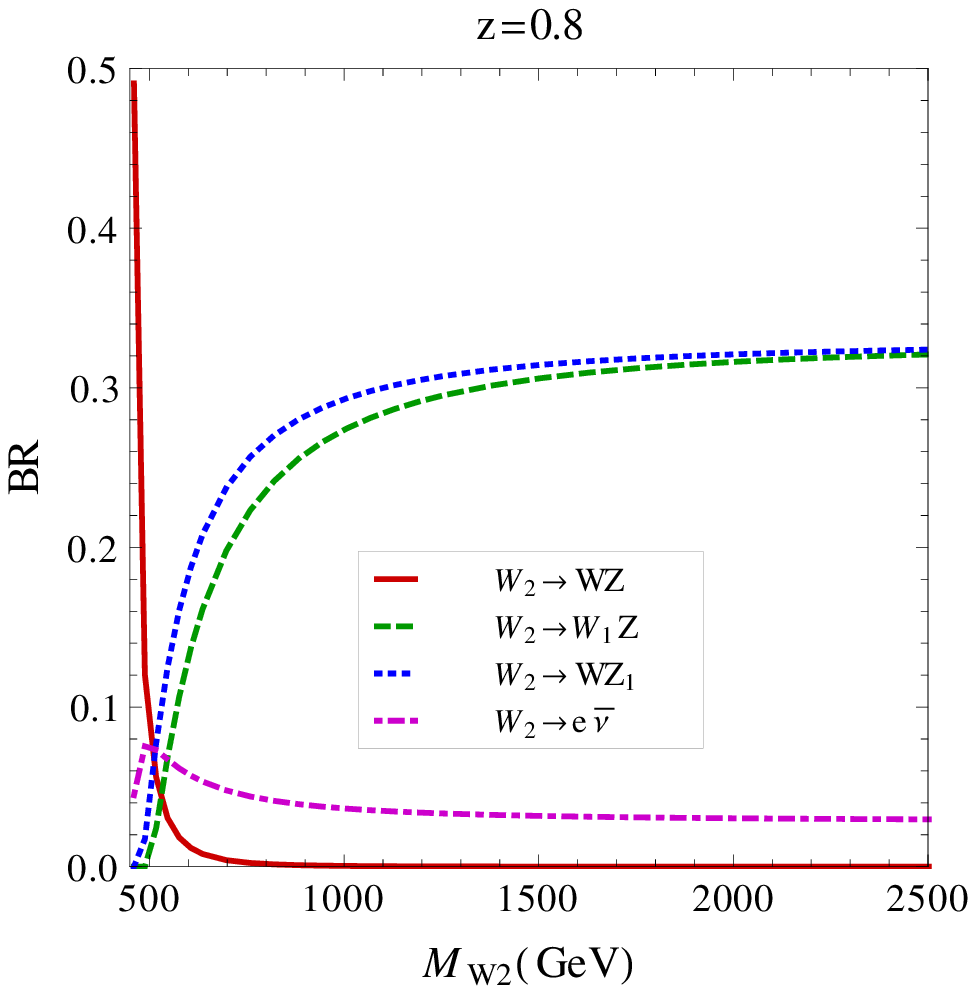,width=7.5cm}}
\put(3.5,-4.8){\epsfig{file=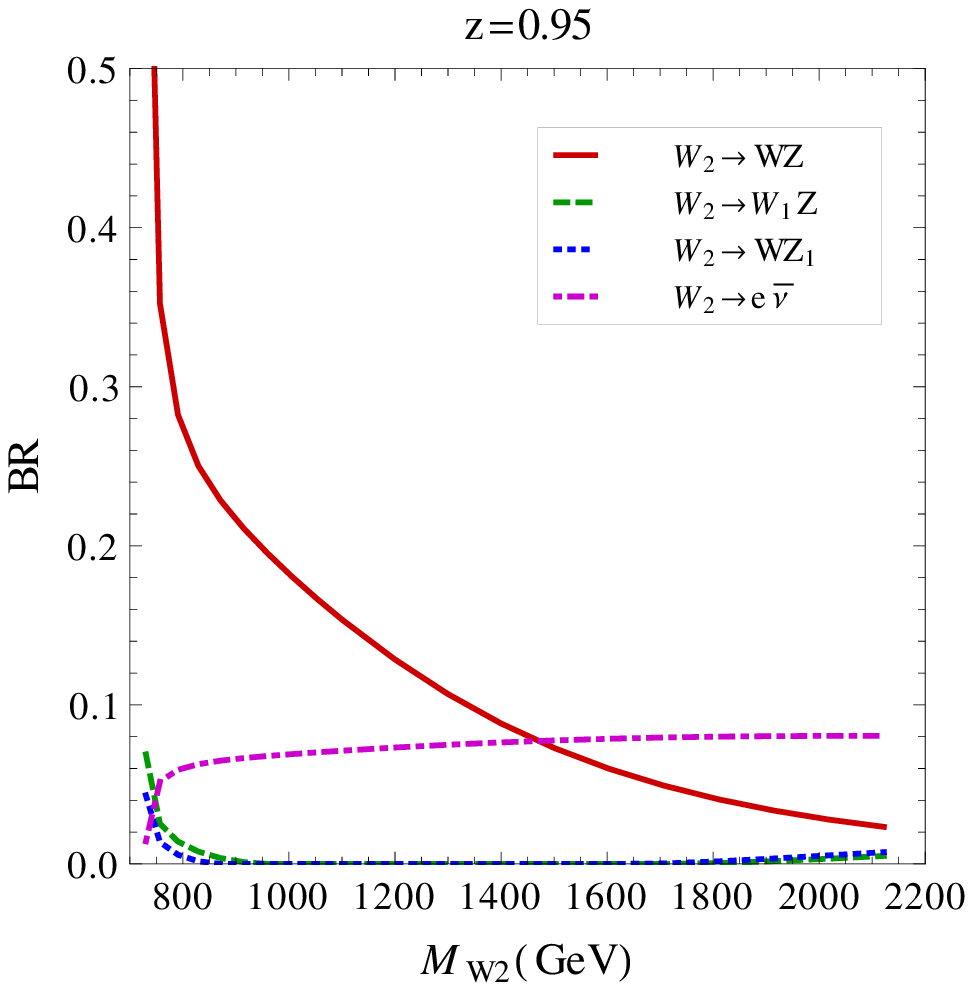,width=7.5cm}}
\end{picture}
\end{center}
\vskip 4.cm
\caption{$W_2$-boson branching ratios as a function of $M_{W2}$ for $z=$0.4,0.6,0.8,0.95.}
\label{fig:BR2}
\end{figure}

Let us now discuss the $W_{1,2}^\pm$-boson branching ratios. The lighter extra gauge bosons, $W_1^\pm$ can decay only into SM 
fermion and gauge boson pairs: $W_1^\pm\rightarrow ff'$ and $W_1^\pm\rightarrow W^\pm Z$. For all $z$-values, the diboson 
channel is the dominant one, the fermion decay being suppressed by EWPT. The decay into electrons, is an increasing function of 
the $z$ parameter, but is always below 2$\%$. However, it can compete with the diboson one if one chooses to rely on clean 
purely leptonic final states as $BR(W_1\rightarrow WZ\rightarrow e\nu_e\mu^-\mu^+)\simeq BR(W_1\rightarrow WZ)/300$. The
 $W_2^\pm$-boson BRs have a more complicated structure. They are displayed in Fig. \ref{fig:BR2} for four representative 
values of the $z$-parameter. The heavier charged extra 
gauge bosons are mainly axial, hence their decay into SM gauge boson pairs is highly suppressed. Their
basic decay modes are into SM fermions and diboson pairs with at least one extra heavy gauge boson: 
$W_2^\pm\rightarrow ff', W_1^\pm Z, W^\pm Z_1, W_1^\pm Z_1$.
As for the lighter extra gauge bosons, the dominant decay mode is the mixed diboson channel 
$W_2^\pm\rightarrow W_1^\pm Z, W^\pm Z_1$. The decay into electron neutrino pairs increases with $z$, never exceeding 10$\%$ 
level. 
But again, if one wants to rely on purely leptonic final states, bosonic and fermionic decay modes compete. For high $z$ 
values, or almost degenerate scenario 
$M_{W1}\simeq M_{W2}$, the latter can even take over owing to the fact that the  
$W_2^\pm\rightarrow W_1^\pm Z, W^\pm Z_1$ modes are kinetically not allowed, as shown in the bottom-right panel of 
Fig. \ref{fig:BR2}.

An exception to this trend appears in the low edge of the spectrum for high $z$ values. In this case, the 
$W_2^\pm\rightarrow W_1^\pm Z, W^\pm Z_1$ channels are kinematically 
suppressed owing to the smallness of 
$\Delta M=M_{W2}-M_{W1}$, and the fermion channel is strongly constrained by EWPT. As a consequence, the only
 decay mode left is $W_2^\pm\rightarrow W^\pm Z$. This channel, a priory sub-dominant owing to the axial nature of 
the $W_2^\pm$-boson, becomes the leading one as shown in the bottom panels of Fig. \ref{fig:BR2},
 giving rise to an extremely narrow resonance.

\section{Drell-Yan production at the LHC and the Tevatron}\label{dy}

Let us now consider the production of the four heavy extra gauge bosons, 
$W_{1,2}^\pm$, predicted by the 4-site Higgsless model at the LHC and the  
Tevatron through the Drell-Yan channel. In contrast with the existing fermiophobic Higgsless 
literature, quite large couplings between SM fermions and extra gauge bosons are
indeed allowed by EWPT as discussed in the previous section.

In the following, we analyze in detail the two charged Drell-Yan processes:
\be
pp\to W,W_{1,2}\to e\nu_e,~~~~~~~ p\bar{p}\to  W,W_{1,2}\to e\nu_e
\ee
at the LHC and the Tevatron, respectively. The two channels differ only by the 
initial state, and are characterized by one isolated electron (or positron) in
the final state plus missing transverse momentum. In our notation, $e\nu_e$ indicates both $e^−\bar\nu_e$ and $e^+\nu_e$. These processes 
can involve the production of 
the four charged extra gauge bosons, $W_{1,2}^\pm$, as intermediate states. They are 
described by the generic formula
\bea \rd\si^{h_1 h_2}(P_1,P_2,p_f) = \sum_{i,j}\int\rd x_1 \rd
x_2~ f_{i,h_1}(x_1,Q^2)f_{j,h_2}(x_2,Q^2)
\,\rd\hat\si^{ij}(x_1P_1,x_2P_2,p_f), \eea 
where $p_f$ summarizes
the final-state momenta, $f_{i,h_1}$ and $f_{j,h_2}$ are the
distribution functions of the partons $i$ and $j$ in the incoming
hadrons $h_1$ and $h_2$ with momenta $P_1$ and $P_2$, respectively,
$Q$ is the factorization scale, and $\hat\si^{ij}$ represent the
cross sections for the partonic processes.
At the LHC, since the two incoming hadrons are protons and we sum over final states with opposite charges, we find
\bea\label{eq:convol_char}
%\refeq{eq:convol}                                  
\rd\si^{h_1h_2}(P_1,P_2,p_f) = \int\rd x_1 \rd x_2 &&\sum_{U=u,c}\sum_{D=d,s}
\Bigl[f_{\bar\PD,\Pp}(x_1,Q^2)f_{\PU,\Pp}(x_2,Q^2)\,\rd\hat\si^{\bar\PD\PU}
(x_1P_1,x_2P_2,p_f)
\nl&&{}
+f_{\bar\PU,\Pp}(x_1,Q^2)f_{\PD,\Pp}(x_2,Q^2)\,\rd\hat\si^{\bar\PU\PD}
(x_1P_1,x_2P_2,p_f)
\nl&&{}
+f_{\PD,\Pp}(x_1,Q^2)f_{\bar\PU,\Pp}(x_2,Q^2)\,\rd\hat\si^{\PD\bar\PU}
(x_1P_1,x_2P_2,p_f)
\nl&&{}
+f_{\PU,\Pp}(x_1,Q^2)f_{\bar\PD,\Pp}(x_2,Q^2)\,\rd\hat\si^{\PU\bar\PD}
(x_1P_1,x_2P_2,p_f)\Bigr].
\eea
Analogously, at the proton-antiproton collider Tevatron we have:
\bea
%\refeq{eq:convol}                                  
\rd\si^{h_1h_2}(P_1,P_2,p_f) = \int\rd x_1 \rd x_2 &&\sum_{U=u,c}\sum_{D=d,s}
\Bigl[f_{\bar\PD,\Pp}(x_1,Q^2)f_{\PU,\bar\Pp}(x_2,Q^2)\,\rd\hat\si^{\bar\PD\PU}
(x_1P_1,x_2P_2,p_f)
\nl&&{}
+f_{\bar\PU,\Pp}(x_1,Q^2)f_{\PD,\bar\Pp}(x_2,Q^2)\,\rd\hat\si^{\bar\PU\PD}
(x_1P_1,x_2P_2,p_f)
\nl&&{}
+f_{\PD,\Pp}(x_1,Q^2)f_{\bar\PU,\bar\Pp}(x_2,Q^2)\,\rd\hat\si^{\PD\bar\PU}
(x_1P_1,x_2P_2,p_f)
\nl&&{}
+f_{\PU,\Pp}(x_1,Q^2)f_{\bar\PD,\bar\Pp}(x_2,Q^2)\,\rd\hat\si^{\PU\bar\PD}
(x_1P_1,x_2P_2,p_f)\Bigr].
\eea

The tree-level amplitudes for the partonic processes have been generated by
means of {\tt PHACT} \cite{Ballestrero:1999md}, a set of routines based on the
helicity-amplitude formalism of \citere{Ballestrero:1994jn}. The matrix
elements have been inserted in the Monte Carlo event generator
{\tt FAST$\_$2f}, dedicated to Drell-Yan processes at the EW and QCD leading
order. {\tt FAST$\_$2f} can compute simultaneously new-physics signal and SM background. It can generate cross-sections and 
distributions for
any observable, including any kind of kinematical cuts. The code is
moreover interfaced with {\tt PYTHIA} \cite{Sjostrand:2006za}. This feature
can allow a more realistic analysis, once {\tt FAST$\_$2f} is matched
with detector simulation programs. 

\subsection{Numerical setup}
\label{se:setup}

For the numerical results presented here, we have used the following
input values \cite{Yao:2006px}: $M_Z=91.187\GeV$, $\GZ=2.512\GeV$, $\GW
=2.105\GeV$,  $\alpha(M_Z)=1/128.88$, $G_F = 1.166\times 10^{-5}$ GeV$^{-2}$.
Additional input parameters are the quark-mixing matrix elements
\cite{Hocker:2001xe}, whose values have been taken to be
$|V_{\Pu\Pd}|=|V_{\Pc\Ps}|=0.975$,
$|V_{\Pu\Ps}|=|V_{\Pc\Pd}|=0.222$, and zero for all other relevant
matrix elements.
In our scheme, the weak mixing-angle and the $W$-boson mass are derived 
quantities. We use  the fixed-width scheme for the matrix element evaluation,
and the CTEQ6L\cite{Pumplin:2002vw} for the parton 
distribution functions at the factorization scale:
\begin{equation}
Q^2={1\over 2}\left (\PT^2(e)+\PT^2(\nu_e )\right )
\end{equation}
where $\PT$ denotes the transverse momentum. This scale choice appears to be 
appropriate for the calculation of differential cross sections, in particular 
for lepton distributions at high energy scales and is adapted from Ref.\cite 
{Dixon:1999di}.
When considering the DY-channel at the LHC, we have moreover implemented the
general set of acceptance cuts defined in  \cite{:2007bs} and here below summarized:
\begin{itemize}
\item {electron (or positron) transverse momentum $\PT(e)>25\GeV$,}
\item {missing transverse momentum $\PTmiss> 25\GeV$,}
\item {electron (or positron) pseudo-rapidity $|\eta_e |< 2.1$,}
\end{itemize}
where $\eta_e=-\log\left (\tan\theta_e/2\right )$, and $\theta_e$ is the polar angle of the charged lepton $e$ with respect 
to the beam. For the analysis at the Tevatron, we include instead a global acceptance of 40$\%$ \cite{Abazov:2010ti}. In both 
cases, we assume 100$\%$ efficiency on charged lepton reconstruction. Additional dedicated kinematical cuts will be described
 in due time. 

For the LHC analysis, we present results at the present center-of-mass 
(c.o.m.) energy 
$\sqrt s=7\TeV$ and two values of the integrated luminosity: $L=1\fba^{-1}$ 
(nowadays) and $L=10\fba^{-1}$ (1 year projection). We moreover give some
hints for the LHC at its design c.o.m. energy $\sqrt s=14\TeV$ and 
$L=10\fba^{-1}$. For the Tevatron, we work within the actual setup:
c.o.m. energy $\sqrt s=1.96\TeV$ and $L=10\fba^{-1}$.

\subsection{$W_{1,2}^\pm$-boson production at the LHC and the Tevatron}
\label{se:boson_production}

\begin{table}[t]
\begin{center}
\begin{tabular}{||c|c|c|c|c||}
\hline
\hline
&$z$&$M_{W1,W2}$(GeV)&$\Gamma_{W1,W2}$(GeV)&$a_{W1,W2}$\\
\hline
a&0.4&410,1000&3.5,24.8&-0.027,0.23\\
\hline
b&0.6&486,794&5.7,15.9&-0.052,0.18\\
\hline
c&0.8&518,636&5.4,2.6&-0.058,0.13\\
\hline
d&0.95&1019,1101&9.4,13.5&-0.062,0.26\\
\hline
\hline
\end{tabular}
\end{center}
\caption{Four representative scenarios for the 4-site Higgsless model.}
\label{tab:cases}
\end{table}
In this section, we analyze the production of the four extra charged gauge bosons in Drell-Yan channel. We consider four 
representative cases of mass spectrum and couplings within the
parameter space allowed by EWPT and unitarity bounds, as shown in Table \ref{tab:cases}.
These four examples give an idea of the possible scenarios predicted
by the 4-site Higgsless model. In the model in fact the ratio
between the masses of the first and second gauge boson triplet, i.e.
$z=M_1/M_2$, is a free parameter. Hence, the distance between the
two masses is arbitrary as well. We have thus
chosen four cases, corresponding to $z$=0.4, 0.6, 0.8, 0.95, and
representing from left to right very distant resonances, the flat-metric
scenario, and a spectrum which tends to degeneracy by increasing $z$.
 
\begin{figure}[t]
\begin{center}
\unitlength1.0cm
\begin{picture}(7,4)
\put(-5.6,-4){\epsfig{file=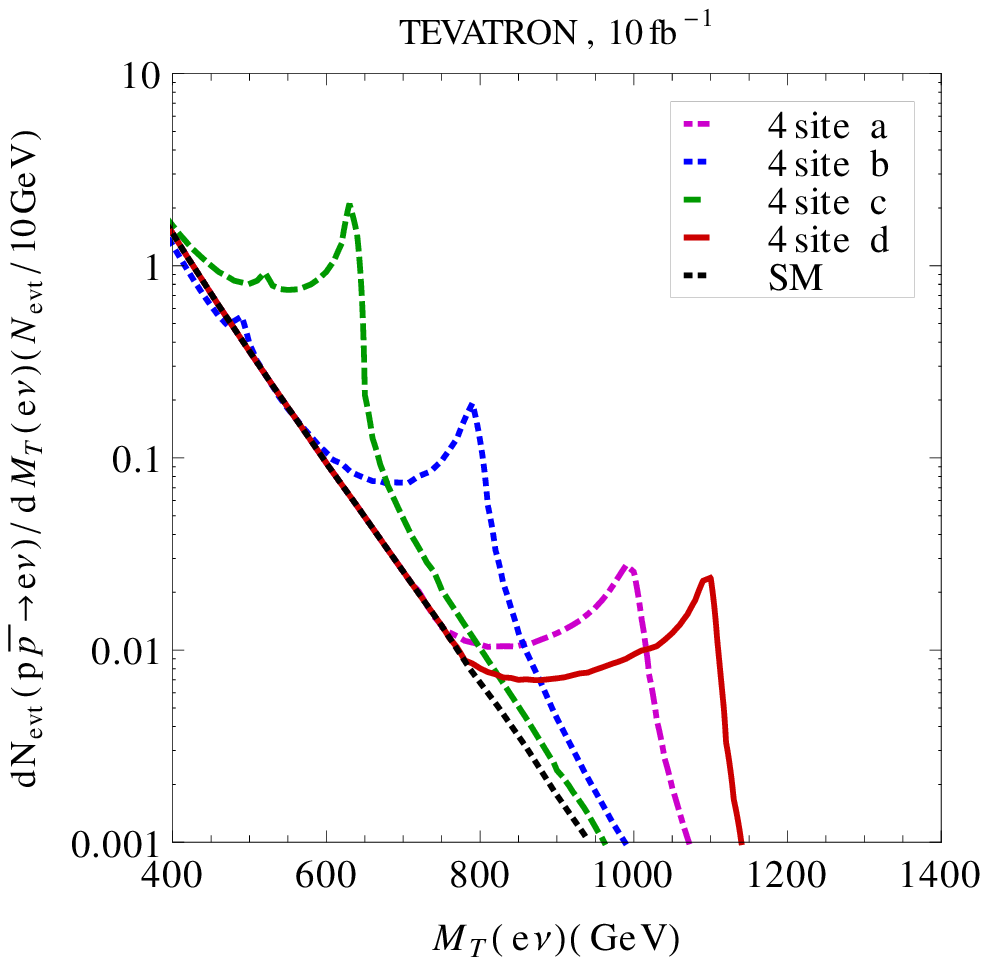,width=7.5cm}}
\put(3.5,-4){\epsfig{file=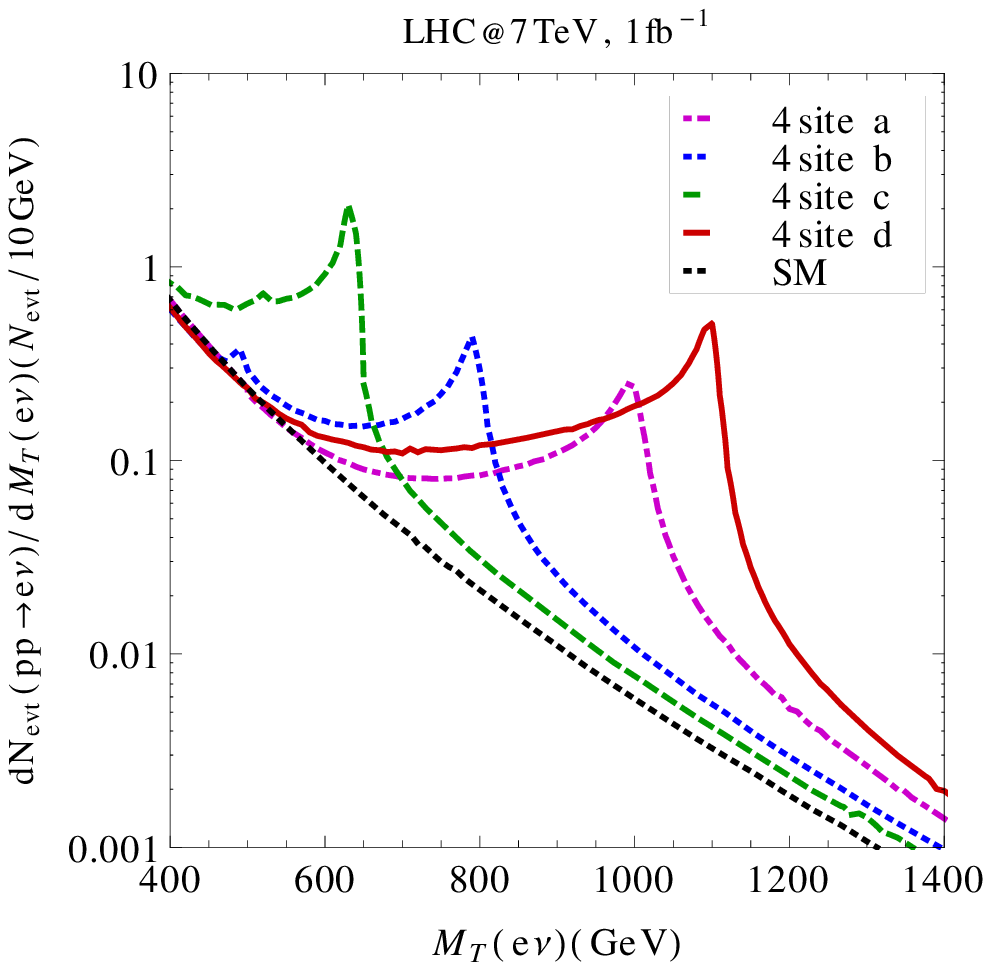,width=7.5cm}}
\end{picture}
\end{center}
\vskip 3.5cm
\caption{Left: Total number of events in a 10$\GeV$-binning versus
the lepton transverse mass, $M_t(e\nu_e)$, for the process
$\Pp\bar\Pp\rightarrow e\nu_e$ at the Tevatron with integrated luminosity L=10 fb$^{-1}$ for the four scenarios of Table
 \ref{tab:cases}. We sum over charge conjugate channels. A global 40$\%$ acceptance is applied. Right: Same distributions 
for the process
$\Pp\Pp\rightarrow e\nu_e$ at the 7 TeV LHC with integrated luminosity L=1 fb$^{-1}$. Standard acceptance cuts are applied.}
\label{fig:Distr_TEV_LHC}
\end{figure}

In order to illustrate spectrum and behaviors of the new heavy $W_{1,2}$ bosons, we have chosen to analyze the distribution in
 the transverse mass of the lepton pair, $M_t(e\nu_e)$,
for the four scenarios of Tab.~\ref{tab:cases}. In Fig.~\ref{fig:Distr_TEV_LHC}, we plot the total number of events as a 
function of the dilepton transverse mass $M_t(e\nu_e)$ at the Tevatron and at the 7 TeV LHC with L=1 fb$^{-1}$. We sum over the
 charged conjugate processes and apply standard acceptance cuts. In most of the cases, one cannot identify the lighter 
resonance. This is a consequence of an intrinsic property of the model. That is, in most part of the parameter space, the 
axial spin-one 
$W_2$-boson is more strongly coupled to fermions than the vector spin-one $W_1$-boson. Thus, the 4-site model appears to 
be degenerate with popular theories predicting a single charged extra gauge boson, $W^\prime$. As a consequence, the same 
experimental analysis performed for the single-resonant benchmark model in \cite{Altarelli:1989ff} could be directly 
applied to the 4-site Higgsless theory. 

In order to estimate the detection rate expected at the Tevatron and the 7 TeV LHC for the
Drell-Yan production of the extra $W_{1,2}^\pm$ gauge bosons, in
Table \ref{tab:evt_TEV_LHC} and we have listed number of SM background events, signal events and corresponding significance 
for the four scenarios given in Tab.~\ref{tab:cases}. We have superimposed the additional dedicated kinematical cut 
$M_t(e\nu_e)\ge M_t^{cut}$, where the value of $M_t^{cut}$ is extracted from Fig. \ref{fig:Distr_TEV_LHC} by taking the point 
where the total number of events intersects the SM background. The Gaussian statistical significance is 
defined as $\sigma=\frac{N_{evt}^T-N_{evt}^B}{a}$, where $N_{evt}^{T(B)}$ is the number of total (background) events and $a$ 
is $\sqrt{N_{evt}^{B}}$ if $N_{evt}^{B}>1$ and 1 otherwise.
 
\begin{table}[t]
\begin{center}
\begin{tabular}{||c|c|c|c|c||c|c|c|c||}
\hline
\hline
&$M_{cut}^{TEV}$ (GeV)&$N_{evt}^{TEV}(B)$&$N_{evt}^{TEV}(T-B)$&$\sigma^{TEV}$&$M_{cut}^{LHC}$ (GeV)&$N_{evt}^{LHC}(B)$&$N_{evt}^{LHC}(T-B)$&$\sigma^{LHC}$\\
\hline
a&751&0.97&3&3&556&18&38&8.9\\
\hline
b&474&38&17&2.8&464&39&45&7.2\\
\hline
c&355&200&159&11.2&327&207&167&11.6\\
\hline
d&781&0.65&2.7&2.7&526&23&78&16.3\\
\hline
\hline
\end{tabular}
\end{center}
\caption{The first column represents the scenario as defined in Tab. \ref{tab:cases}. The following first four columns give
 the lower bound on the dilepton transverse mass, the number of expected background events, the number of signal events and 
the corresponding statistical significance at the Tevatron with 10 fb$^{-1}$. The remaining four columns contain the same data
 for the LHC with 1 fb$^{-1}$.}
\label{tab:evt_TEV_LHC}
\end{table}

\begin{figure}[t]
\begin{center}
\unitlength1.0cm
\begin{picture}(7,4)
\put(-5.6,-4){\epsfig{file=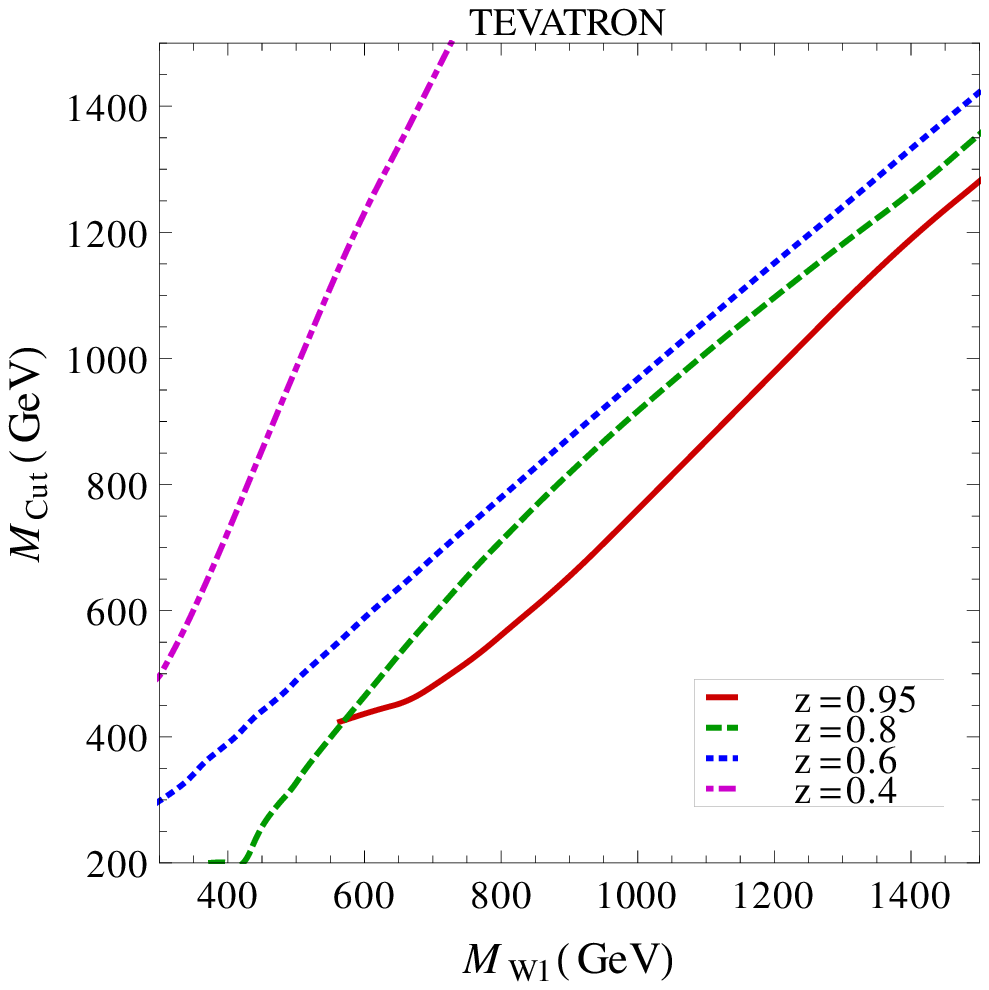,width=7.5cm}}
\put(3.5,-4){\epsfig{file=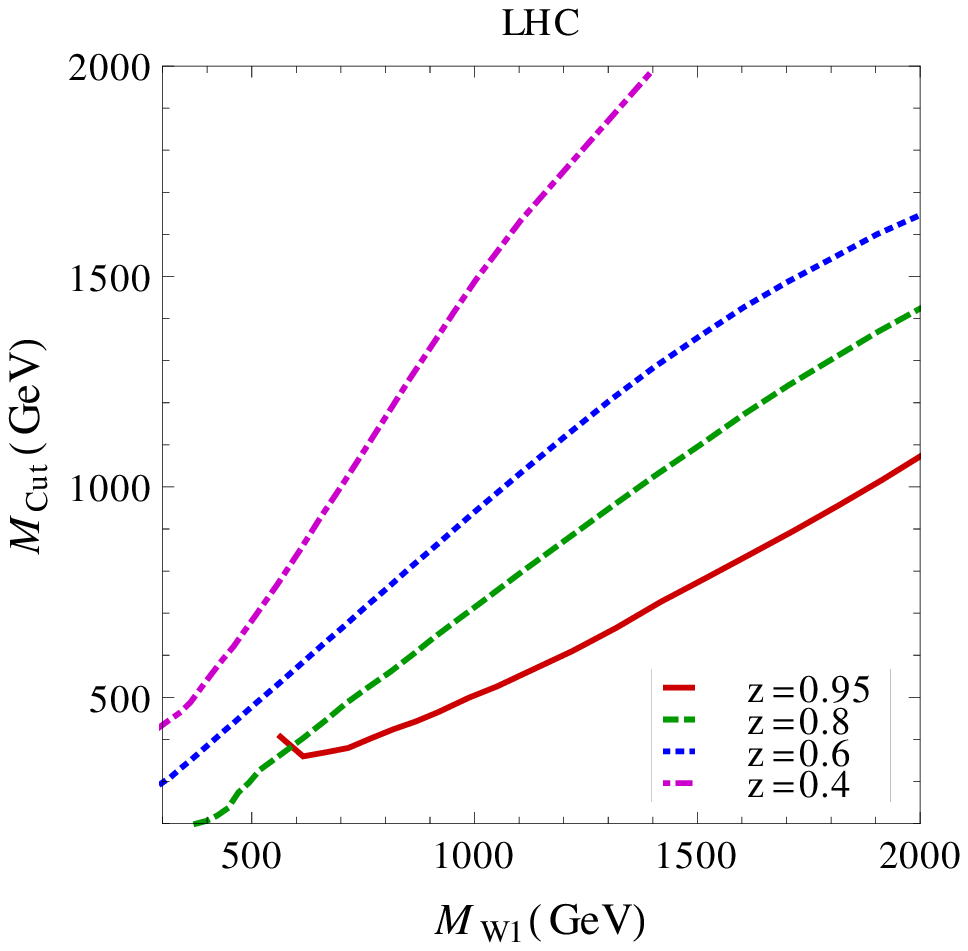,width=7.5cm}}
\end{picture}
\end{center}
\vskip 3.5cm
\caption{Lower bound on the dilepton transverse mass, $M_{cut}$, as a function of the lighter extra gauge boson mass, 
$M_{W1}$, for four values of the free $z$-parameter: $z$=0.4, 0.6, 0.8, 0.95. At Tevatron (left) and LHC (right)}
\label{fig:Mcut_LHC}
\end{figure}

In order to enhance the signal-over-background ratio, the lower bound on the dilepton transverse mass, $M_{cut}$, must be
 indeed chosen \textit{ad hoc}. In Fig.~\ref{fig:Mcut_LHC}, we show how $M_{cut}$ 
varies with $M_{W1}$ for the four representative $z$-values in Tab. \ref{tab:cases}. 
Left and right panels of Fig.~\ref{fig:Mcut_LHC} refer to the Tevatron and the 7 TeV LHC, respectively. We note that $M_{cut}> M_{W1}$ for $z\leq0.5$. This 
means that the lighter resonances are in a region where the 4-site model 
predicts a depletion of events compared to the SM.

\begin{figure}[t]
\begin{center}
\unitlength1.0cm
\begin{picture}(7,10)
\put(-5.6,2.7){\epsfig{file=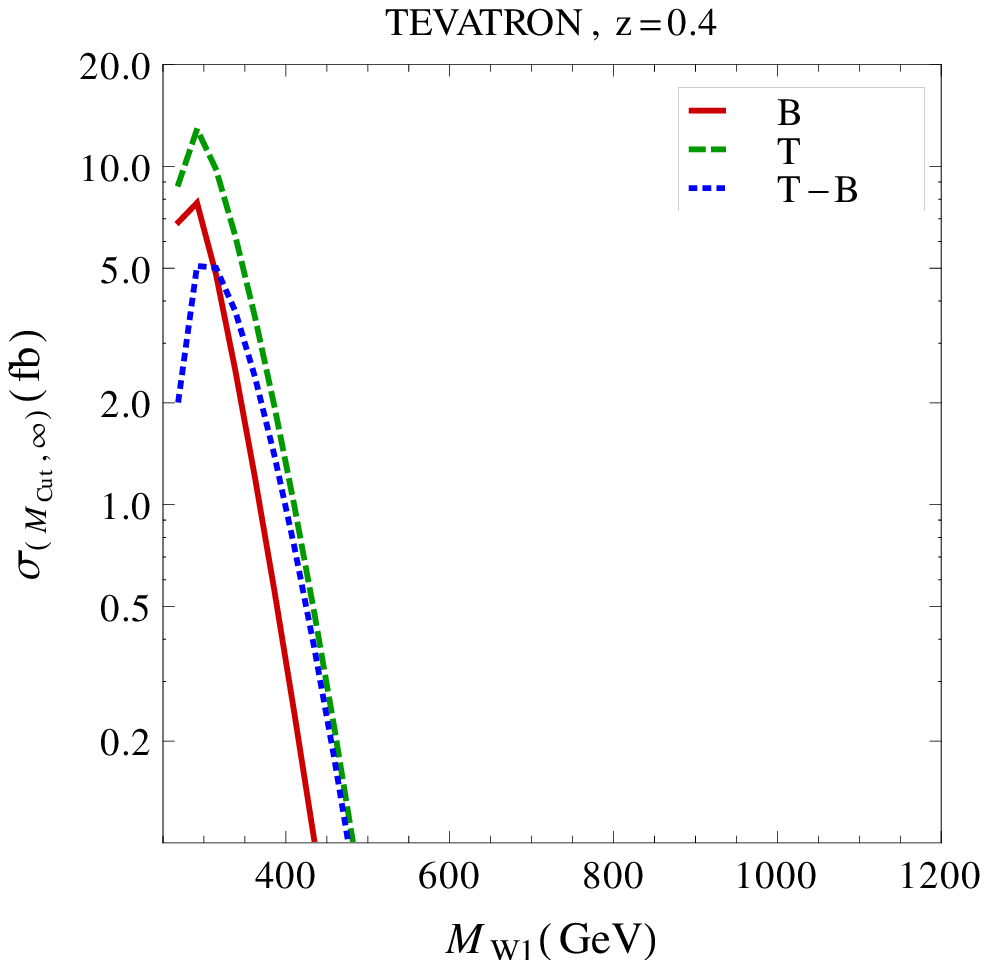,width=7.5cm}}
\put(3.5,2.7){\epsfig{file=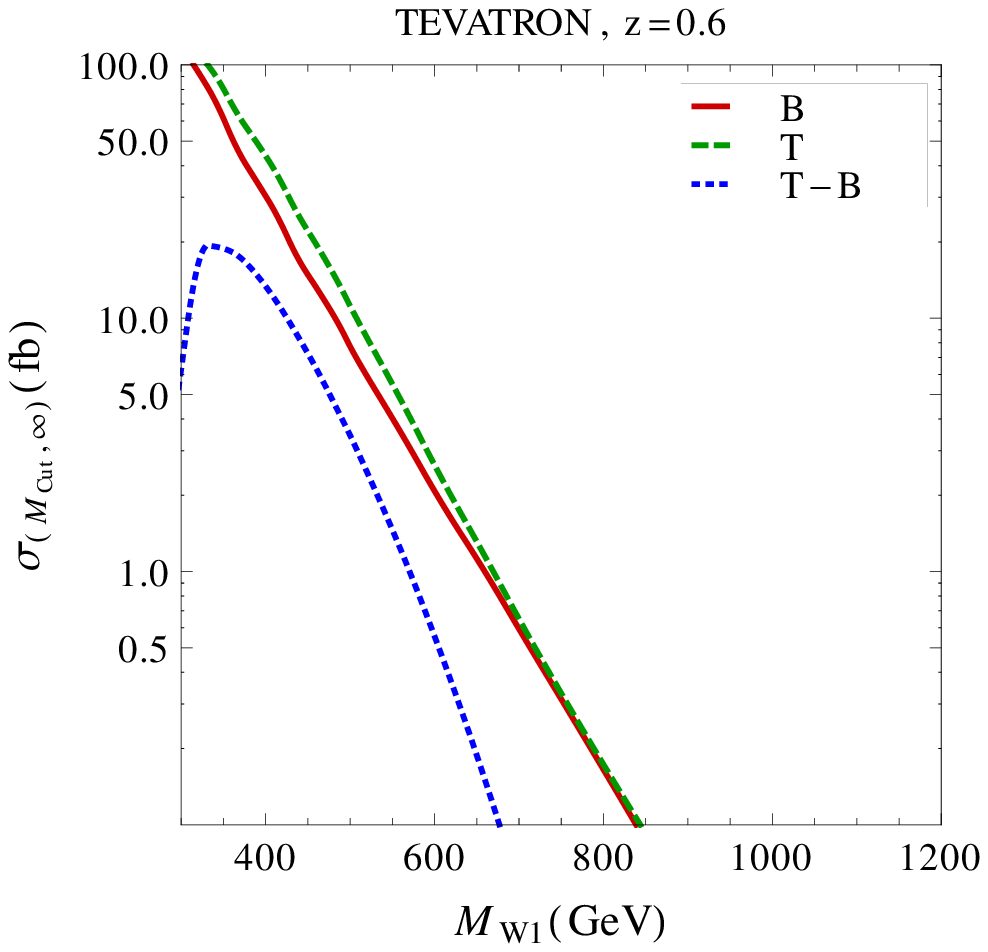,width=7.5cm}}
\put(-5.6,-4.8){\epsfig{file=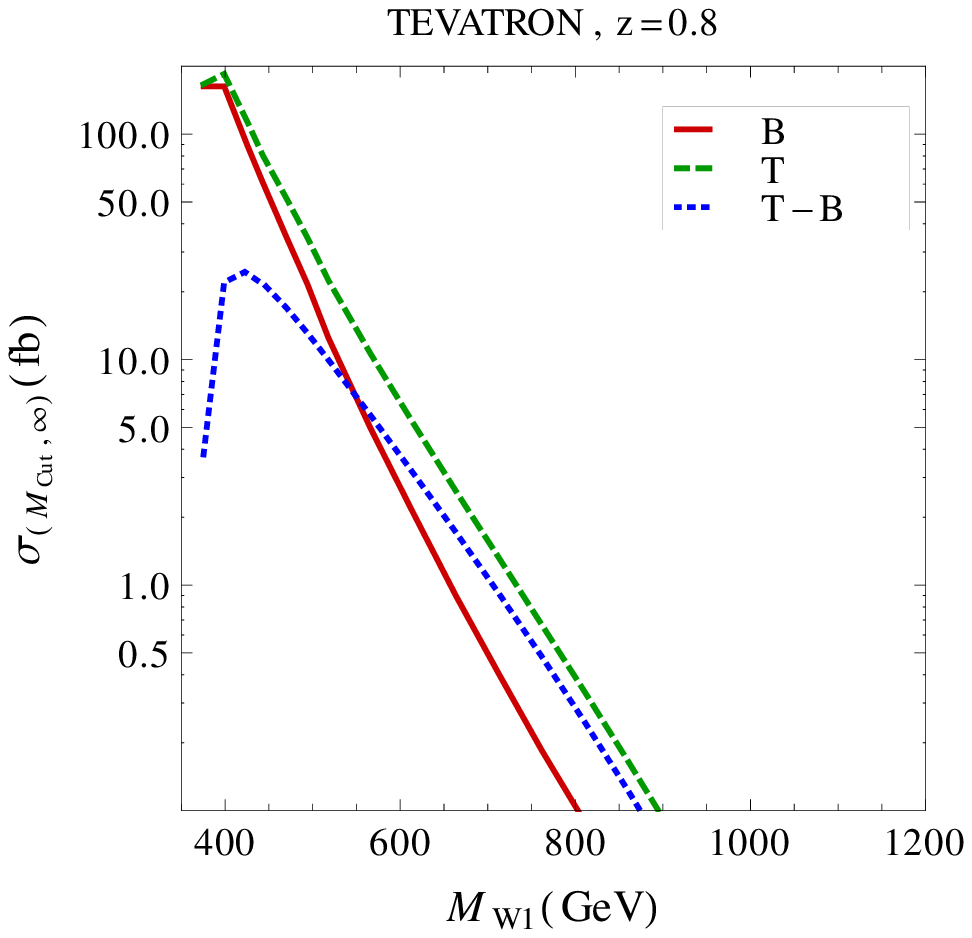,width=7.5cm}}
\put(3.5,-4.8){\epsfig{file=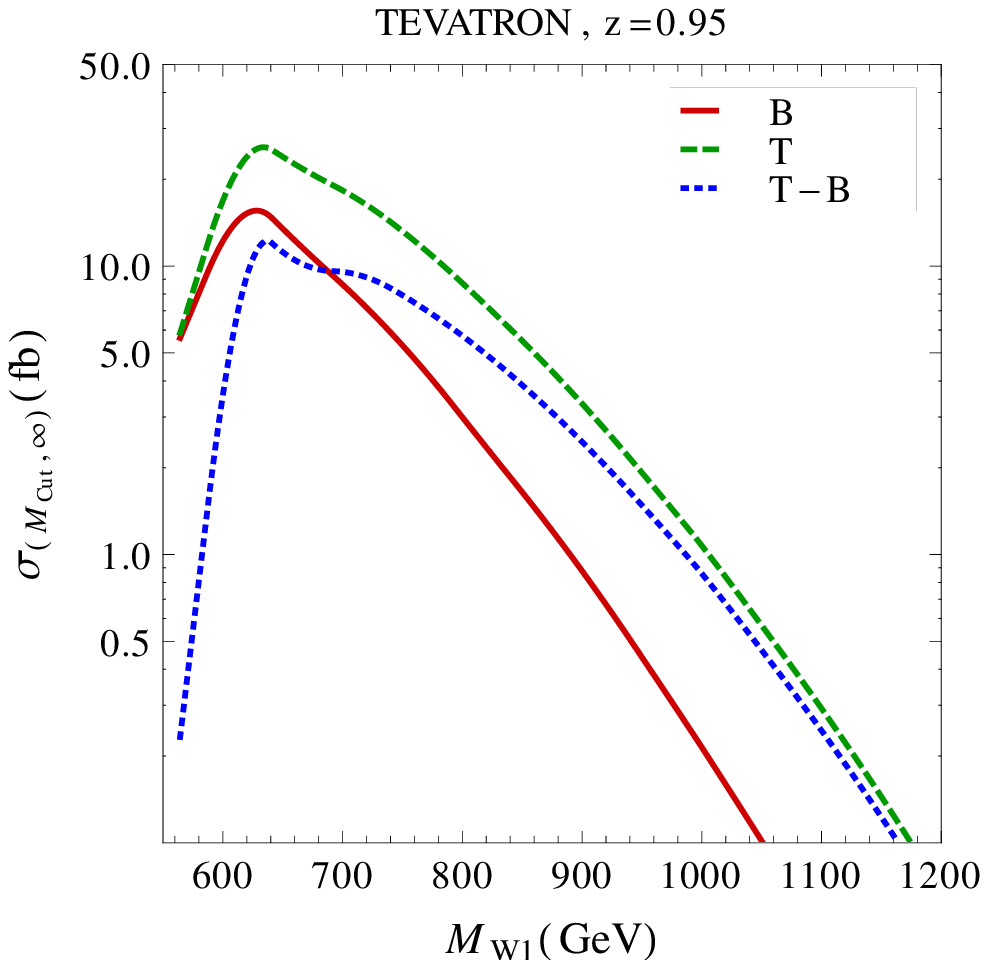,width=7.5cm}}
\end{picture}
\end{center}
\vskip 4.cm
\caption{$W_{1,2}^\pm$-boson cross-sections, for the maximal $a_{W1}$ coupling allowed by
EWPT, integrated over the mass window $M_T(e\nu_e)\ge M_{cut}$, as function of $M_{W1}$ at the Tevatron. The dashed line
 represents the total cross-section (T), including
the interference between $W_{1,2}$-boson signal and SM background. The solid line shows the SM background (B). The dotted 
line corresponds to their difference: S=T-B.  
In the four plots, the free $z$-parameter assumes the values: $z$=0.4, 0.6, 0.8, 0.95.}
\label{fig:C-S_TEV_excl}
\end{figure}

\begin{figure}[!htbp]
\begin{center}
\unitlength1.0cm
\begin{picture}(7,10)
\put(-5.6,2.7){\epsfig{file=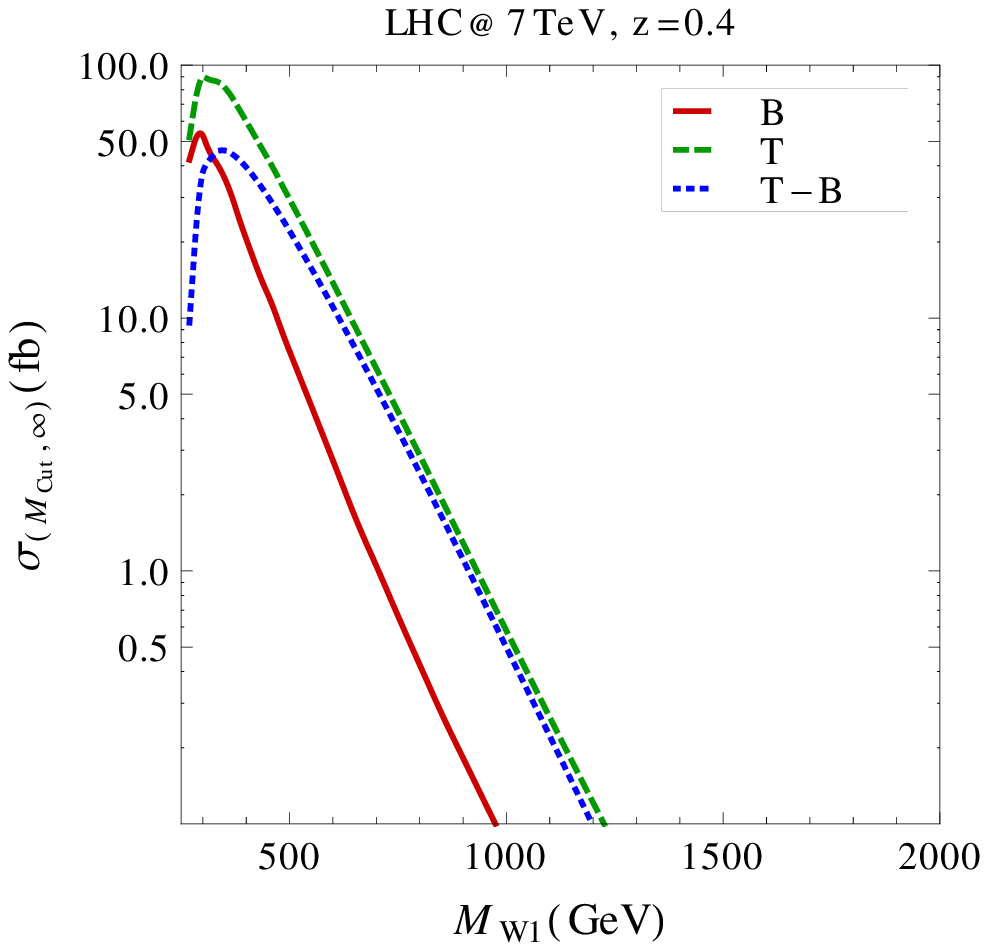,width=7.5cm}}
\put(3.5,2.7){\epsfig{file=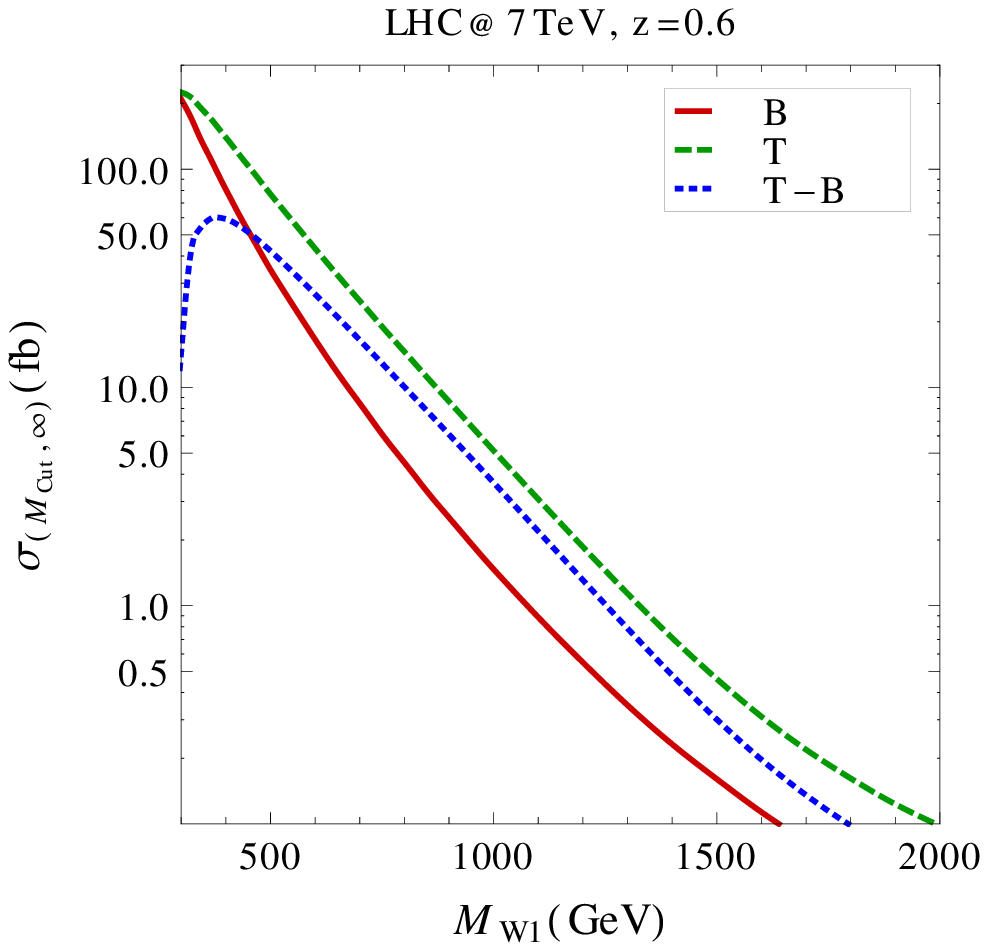,width=7.5cm}}
\put(-5.6,-4.8){\epsfig{file=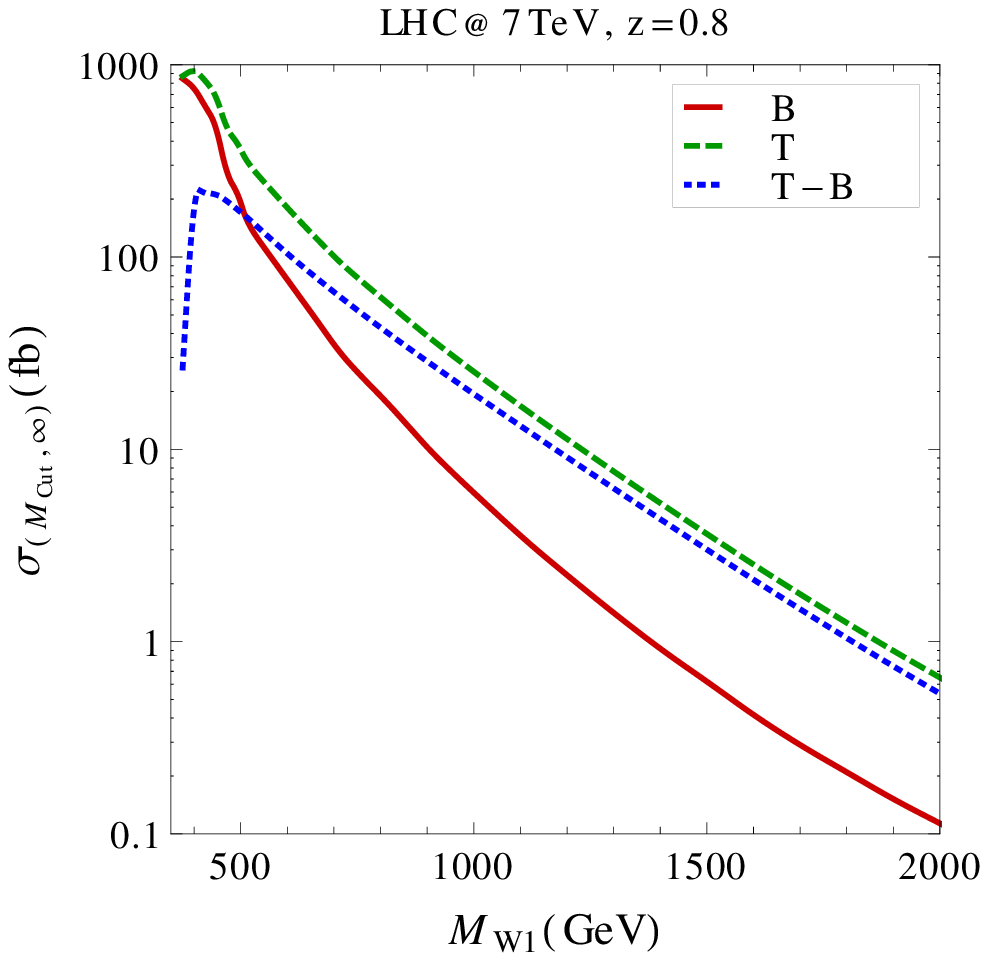,width=7.5cm}}
\put(3.5,-4.8){\epsfig{file=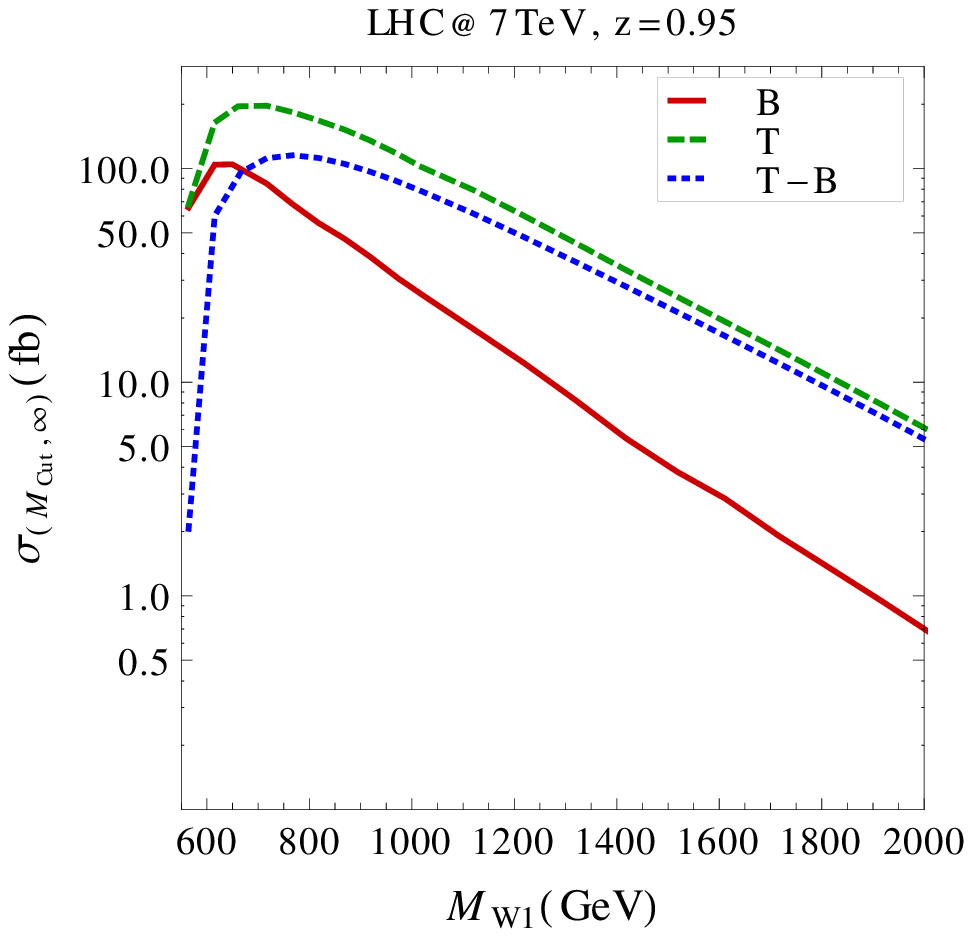,width=7.5cm}}
\end{picture}
\end{center}
\vskip 4.cm
\caption{$W_{1,2}^\pm$-boson cross-sections, for the maximal $a_{W1}$ coupling allowed by
EWPT, integrated over the mass window $M_T(e\nu_e)\ge M_{cut}$, as function of $M_{W1}$ at the 7 TeV LHC. The dashed line
 represents the total cross-section (T), including
the interference between $W_{1,2}$-boson signal and SM background. The solid line shows the SM background (B). The dotted
 line corresponds to their difference: S=T-B.  
In the four plots, the free $z$-parameter assumes the values: $z$=0.4, 0.6, 0.8, 0.95.}
\label{fig:C-S_LHC_excl}
\end{figure}

The corresponding total cross-sections integrated over the window, 
$M_T(e\nu_e)\ge M_{cut}$,
are shown in Fig.~\ref{fig:C-S_TEV_excl} for the Tevatron and Fig.~\ref{fig:C-S_LHC_excl} for the 
7 TeV LHC as a function of $M_{W1}$. We consider the usual four $z$-values: $z$=0.4, 0.6, 0.8, 0.95.
These are just bare values, useful only to give an idea of the magnitude of the expected cross-sections around the
 resonances. The displayed cross-sections have been indeed calculated for the maximum value of the $W_1$-boson coupling to
 electrons, $a_{W1}$, allowed by EWPT at fixed $z$.

At fixed mass, $M_{W1}$, the cross-section gets larger by increasing the value of the
$z$-parameter. This effect is due to the fact that high $z$-values help in relaxing the EWPT constraints on the $W_{1,2}$-boson
 coupling to SM fermions, as previously shown in Fig. 
\ref{fig:EWPT_a1-a2}. The signal cross section at the Tevatron can thus lie in the range:
5 fb $\le\sigma_{T-B}\le$ 25 fb for the $W_1$-boson mass interval allowed by EWPT and unitarity. 
At the 7 TeV LHC, the expected cross section increases by about an order of magnitude, reaching values of a few hundred fb.

%In Fig.\ref{fig:Sig_LHC} we plot the statistical significance for the maximal $a_{W_1}$ allowed by
%EWPT, integrated over the mass window given in Eq.(\ref{window}) as function of $M_{W_1}$ and %$z$.
%\begin{figure}[!htbp]
%\begin{center}
%\unitlength1.0cm
%\begin{picture}(7,4)
%\put(-5.6,-4){\epsfig{file=Signif_TEV10_Excl.eps,width=7.5cm}}
%\put(3.5,-4){\epsfig{file=Signif_LHC_Excl.eps,width=7.5cm}}
%\end{picture}
%\end{center}
%\vskip 3.5cm
%\caption{Left: Statistical significance, defined as in Eq.~\ref{sig}, for the maximal $a_{W_1}$ %allowed by
%EWPT, integrated over the mass window given in Eq.(\ref{window})-exclusion as function of %$M_{W_1}$ and $z$, at Tevatron
%(top) and LHC (bottom). Right: the same for discovery}
%\label{fig:Sig_LHC}
%\end{figure}

\section{$W_{1,2}^{\pm}$ exclusion and discovery reach at the Tevatron and LHC}
\label{disco}

In this section, we discuss the prospects of discovering the four charged 
spin-1 bosons predicted by the 4-site Higgsless model at the LHC.
Let us start by deriving the present exclusion limits on the $W_{1,2}$-bosons 
from the Tevatron experiment. We consider both neutral and charged Drell-Yan 
channels, $p\bar p\to e^-e^+$ and $p\bar p\to e\nu_e$ , at the collected 
luminosity L=10 fb$^{-1}$, and we include a global 40$\%$ acceptance and  
efficiency factor. For the neutral Drell-Yan channel, we compute the expected 
number of events in the asymmetrical mass window 
$M_{inv}(e^+e^−)\ge M_{Z1} - 3R_{TEV}$, where  $R_{TEV}\simeq 3.4\%M$ is the 
approximated D0 mass resolution (see [47] and references therein).
For the charged Drell-Yan channel instead, we consider the mass 
window: $M_t(e\nu_e)\ge M_{cut}$ as previously discussed. For both processes, 
we then evaluate the region of the parameter space where the Gaussian 
statistical significance is bigger than 2: 
$\sigma=\frac{N_{evt}^T-N_{evt}^B}{a}\geq2$, where $N_{evt}^{T(B)}$ is the 
number of total (SM background) events and $a=\sqrt{N_{evt}^{B}}$ if 
$N_{evt}^{B}>1$ and $a=1$ otherwise.
\begin{figure}[t]
\begin{center}
\unitlength1.0cm
\begin{picture}(7,10)
\put(-5.6,2.7){\epsfig{file=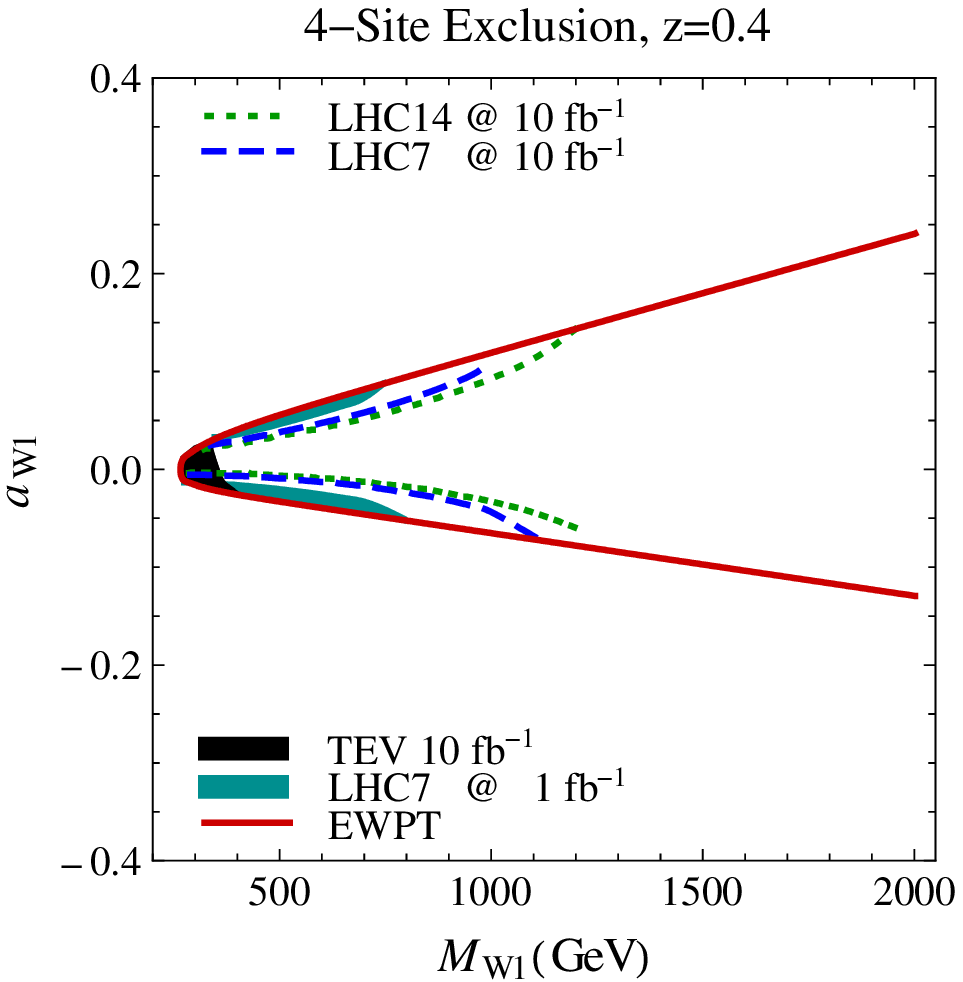,width=7.5cm}}
\put(3.5,2.7){\epsfig{file=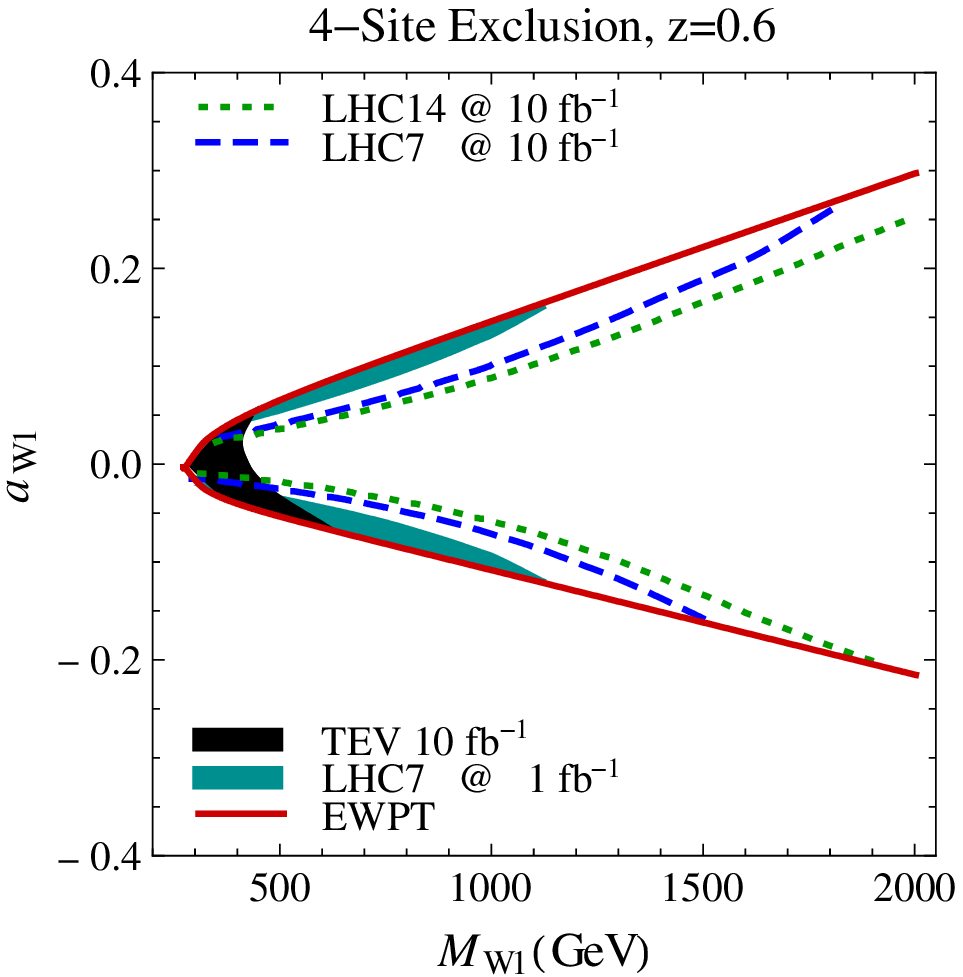,width=7.5cm}}
\put(-5.6,-5){\epsfig{file=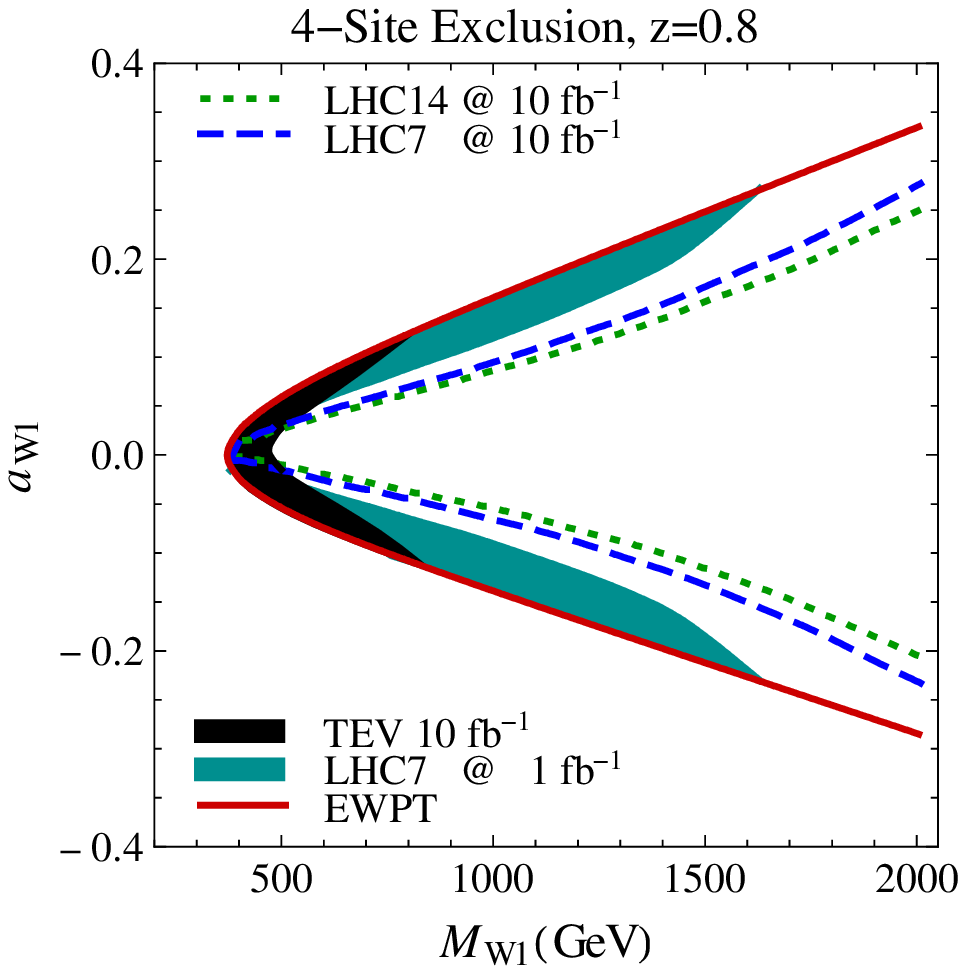,width=7.5cm}}
\put(3.5,-5){\epsfig{file=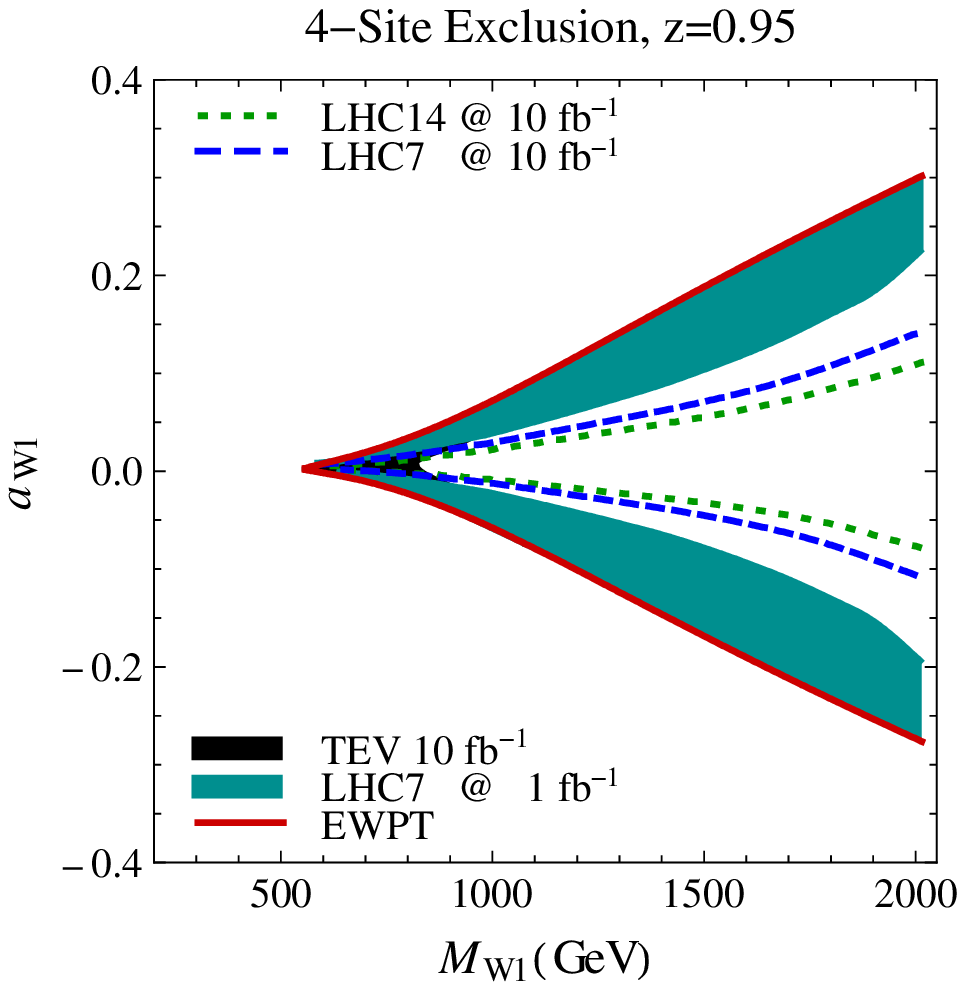,width=7.5cm}}
\end{picture}
\end{center}
\vskip 4.cm
\caption{95$\%$ CL exclusion limits in the $(M_{W1},a_{W1})$ plane at the 
7 TeV LHC with an integrated luminosity of 10 fb$^{-1}$ (blue dashed line), 
and at the 14 TeV LHC with 10 fb$^{-1}$ (dotted green line). The red solid 
contour defines the parameter space allowed by EWPT and unitarity. The black 
shaded region represents the 95$\%$ CL exclusion limits from direct 
searches in the Drell-Yan channel at the Tevatron with L=10 fb$^{-1}$. 
The cyan shaded region gives the expected 95$\%$ CL exclusion limits 
at the actual 7 TeV LHC with L=1 fb$^{-1}$. From top-left to bottom-right, 
the $z$-parameter is fixed to be: $z=$ 0.4, 0.6, 0.8 and 0.95.}
\label{fig:E_LHC}
\end{figure}
The expected 95$\%$ CL exclusion region in the plane $M_{W1}, a_{W1}$ is 
depicted in Fig. \ref{fig:E_LHC} by the black shaded area. The 4-site 
model can be  constrained by present data like other popular $W^\prime$ 
theories. Assuming maximal values for the $W_{1,2}$-boson couplings to SM 
fermions, the direct limit on the mass of the heavy extra resonances is 
indeed $M_{W1}\ge$ 1100 GeV for $z$=0.95 as displayed in the bottom-right 
panel of Fig. \ref{fig:E_LHC}. 
%For smaller $z$-values, EWPT give very strong 
%bounds on the $a_{W1,W2}$ couplings so that the Tevatron is less and less 
%effective. 
For $z$=0.4 on the other hand, the mass bound is about $M_{W1}\ge$ 400 GeV 
as shown in the top-left panel of Fig.\ref{fig:E_LHC}. This is not surprising
as the corresponding bound on $M_{W2}$ would be 1 TeV as above, and the 
lighter resonance would actually be invisible. These limits mainly come from 
the neutral Drell-Yan channel, which might involve the production and decay 
of the two neutral extra gauge bosons, $Z_{1,2}$, predicted by the 4-site 
model (see \cite{Accomando:2010ir,Accomando:2008jh,Accomando:2008dm}). Here, 
they have been appropriately converted in order to appear in the parameter 
space expressed in terms of the charged $M_{W1}, a_{W1}$ physical observables. 

In the same Fig.\ref{fig:E_LHC}, we also show the 95$\%$ CL exclusion limits 
that one could derive at different stages of the LHC. These contour plots 
have been computed by integrating the cross section in the domain 
$M_t(e\nu_e)\ge M_{cut}$, and assuming the acceptance times efficiency setup 
described in Sec. \ref{se:setup}.
The cyan shaded area in Fig.\ref{fig:E_LHC} shows the 95$\%$ CL exclusion 
limits on $W_{1,2}$-boson mass and coupling that one could extract from 
present data at the 7 TeV LHC, which has now collected over 1 fb$^{-1}$. As 
one can see the LHC greatly extends the Tevatron bounds, 
being able to exclude the entire mass spectrum in the almost degenerate 
scenario not only for maximal couplings. 
We furthermore compare this result with the 95$\%$ CL exclusion limit one 
could reach in one year from now, that is at the 7 TeV LHC with an expected 
integrated luminosity L=10 fb$^{-1}$. In this second stage, the LHC could 
exclude the 4-site model up to energy scales of the order of $M_{W1}\ge$ 
1100, 1800, 2000 GeV for maximal $a_{W1,W2}$ couplings allowed by EWPT and 
$z$=0.4, 0.6, 0.8 (or 0.95) respectively. Finally, we analyze the energy 
dependence of such limits, displaying the 95$\%$ CL exclusion limits at the 
project 14 TeV LHC with L=10 fb$^{-1}$. This upgrade would further extend the 
exclusion potential by a few hundreds of GeV. For high-intermediate values of 
the free $z$-parameter, a sizeable portion of the parameter space could be 
analyzed and perhaps excluded. The low $z$ range would need higher energy and 
luminosity.   

\begin{figure}[t]
\begin{center}
\unitlength1.0cm
\begin{picture}(7,10)
\put(-5.6,2.7){\epsfig{file=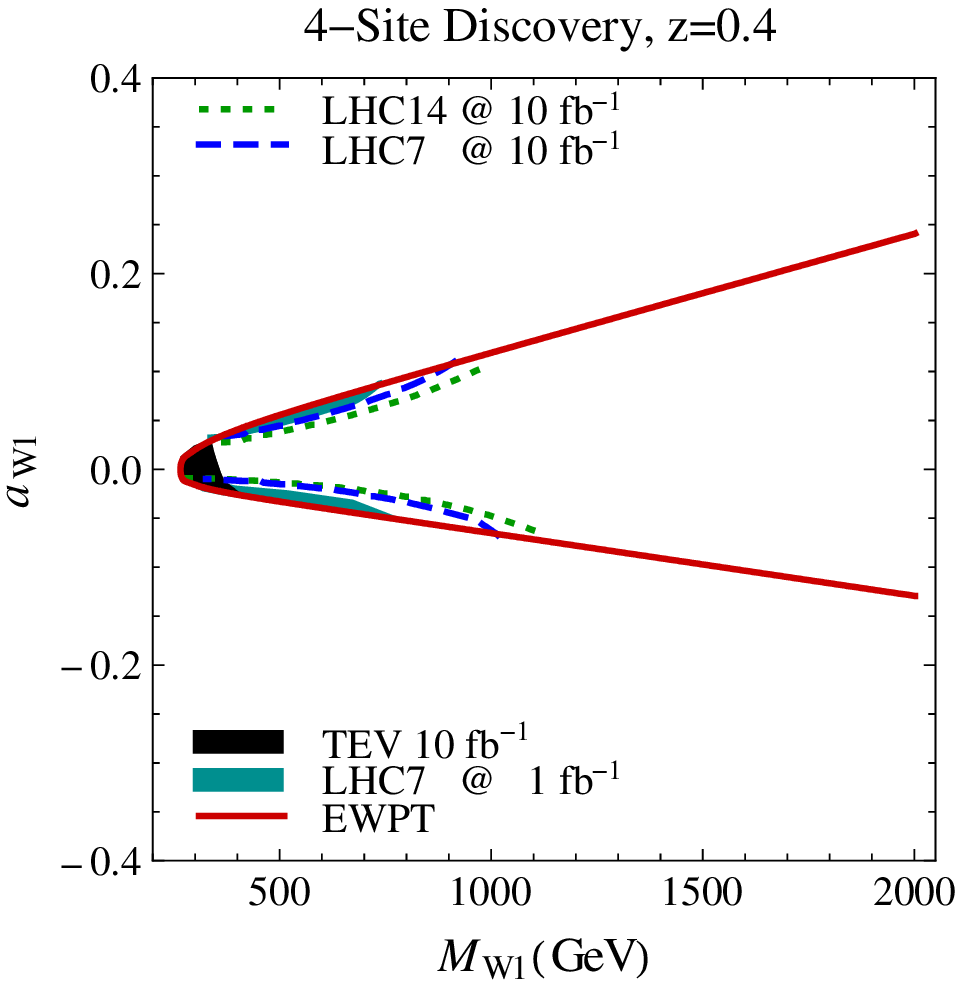,width=7.5cm}}
\put(3.5,2.7){\epsfig{file=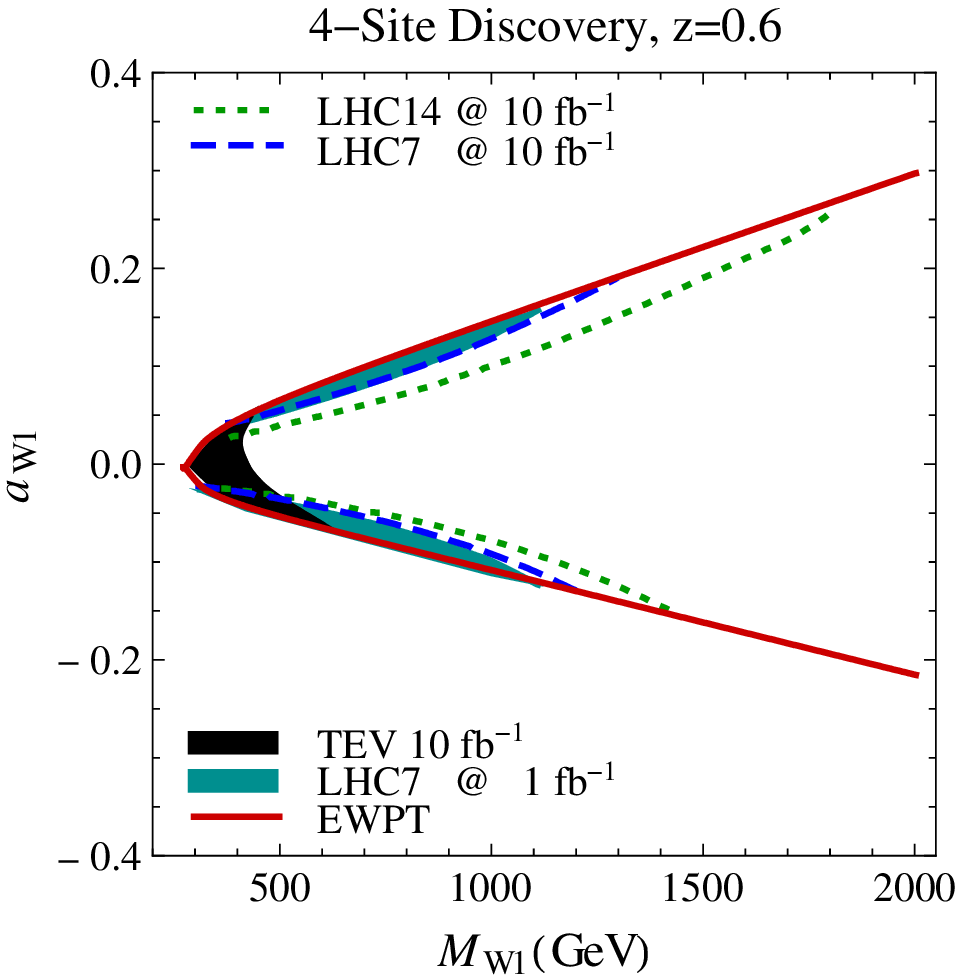,width=7.5cm}}
\put(-5.6,-5){\epsfig{file=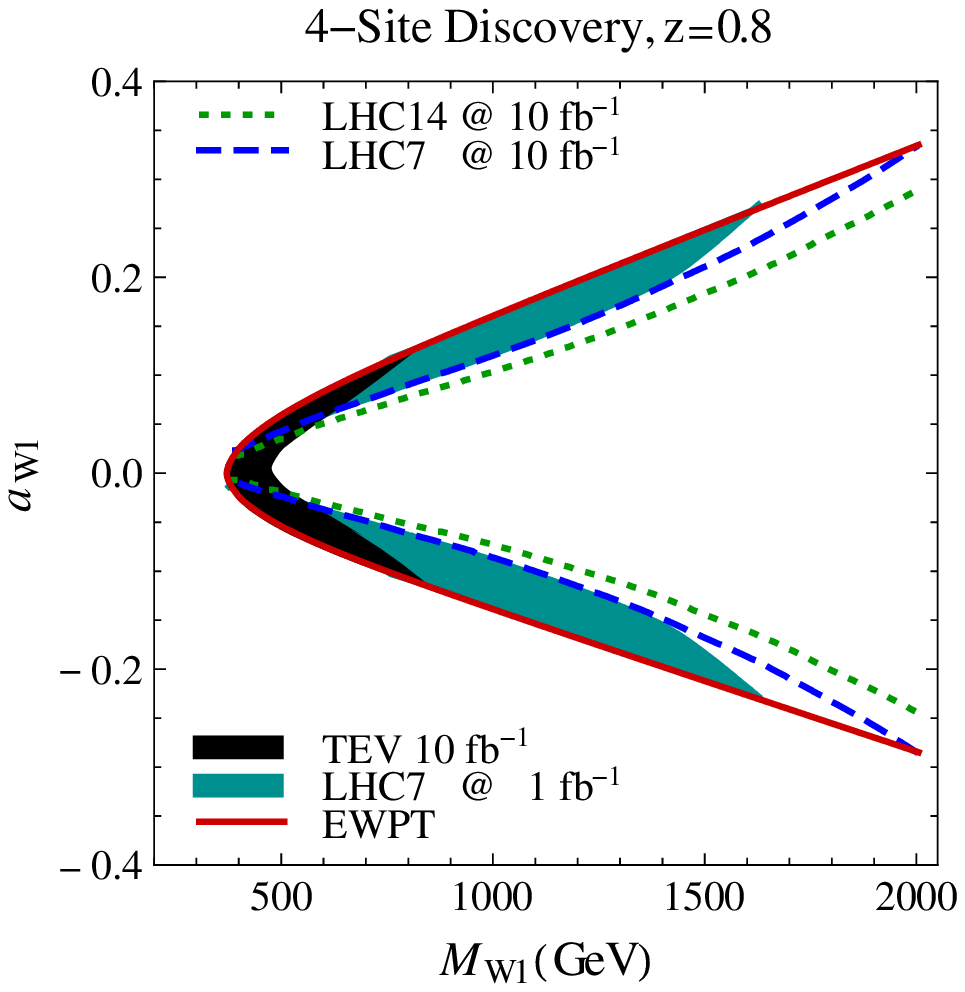,width=7.5cm}}
\put(3.5,-5){\epsfig{file=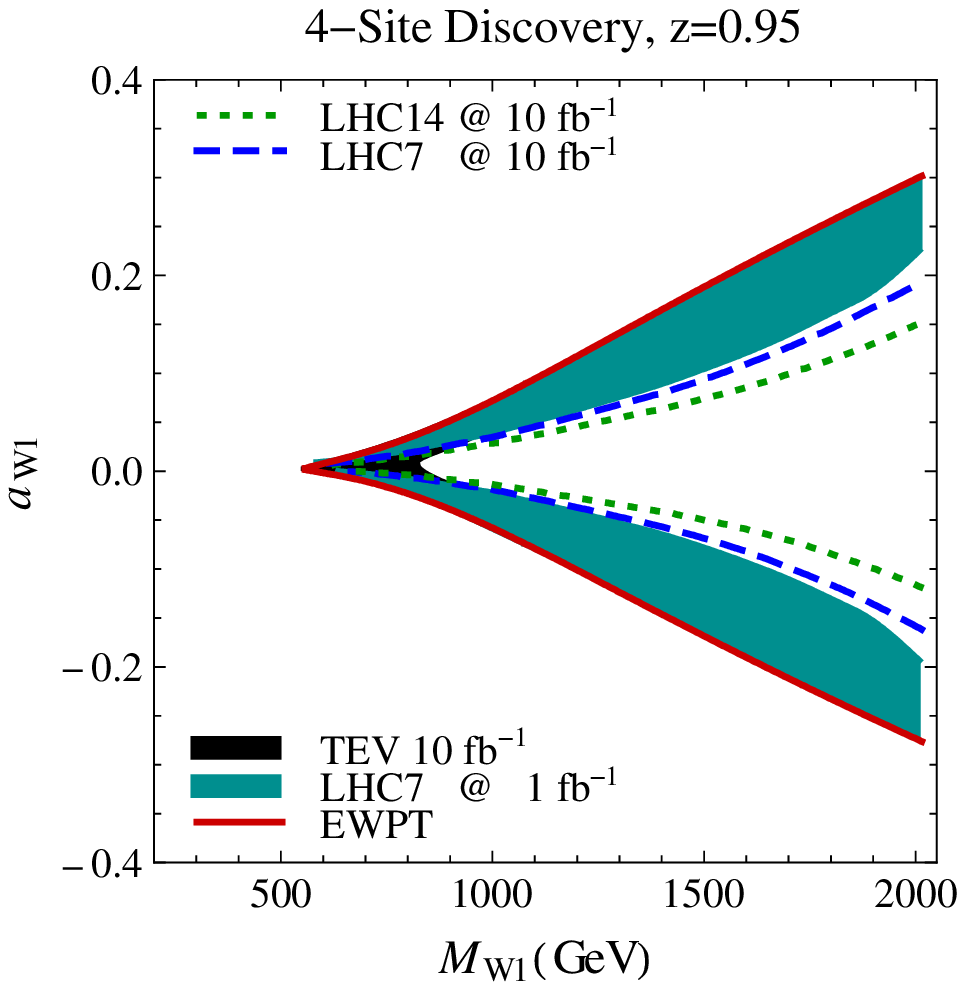,width=7.5cm}}
\end{picture}
\end{center}
\vskip 4.cm
\caption{$W_{1,2}$-boson discovery in the $(M_{W1},a_{W1})$ plane at the 
7 TeV LHC with an integrated luminosity L=10 fb$^{-1}$ (blue dashed line), 
and at the 14 TeV LHC with 10 fb$^{-1}$ (dotted green line). The red solid 
contour defines the parameter space allowed by EWPT and unitarity. 
The cyan shaded region gives the expected 95$\%$ CL exclusion limits
at the actual 7 TeV LHC with L=1 fb$^{-1}$.
From top-left to bottom-right, the $z$-parameter is fixed to be: 
$z=$ 0.4, 0.6, 0.8 and 0.95.}
\label{fig:D_LHC}
\end{figure}

In Fig. \ref{fig:D_LHC}, we show the prospects of discovering the four 
charged spin-1 bosons predicted by the 4-site Higgsless model during the 
LHC early and future stages, as done above for the exclusion (now simply 
requiring $\sigma\ge$ 5).

The dashed blue line gives the 5$\sigma$-discovery potential in the next 
one-year period at the 7 TeV LHC with L=10~fb$^{-1}$. The dotted green line 
projects the discovery reach at the 14 TeV LHC with L=10~fb$^{-1}$. As for 
the exclusion, the discovery contour plots have been obtained by integrating 
the cross section over the domain $M_t(e\nu_e)\ge M_{cut}$, and assuming the 
acceptance times efficiency setup described in Sec. \ref{se:setup}. The 
global discovery reach during the early run of the LHC is substantial. The
charged extra gauge bosons could be detected up to $M_{W1}\sim$ 1000, 1300, 
2000 GeV for $z$ = 0.4, 0.6, 0.8 (or 0.95) respectively.

\begin{figure}[t]
\begin{center}
\unitlength1.0cm
\begin{picture}(7,10)
\put(-5.6,2.7){\epsfig{file=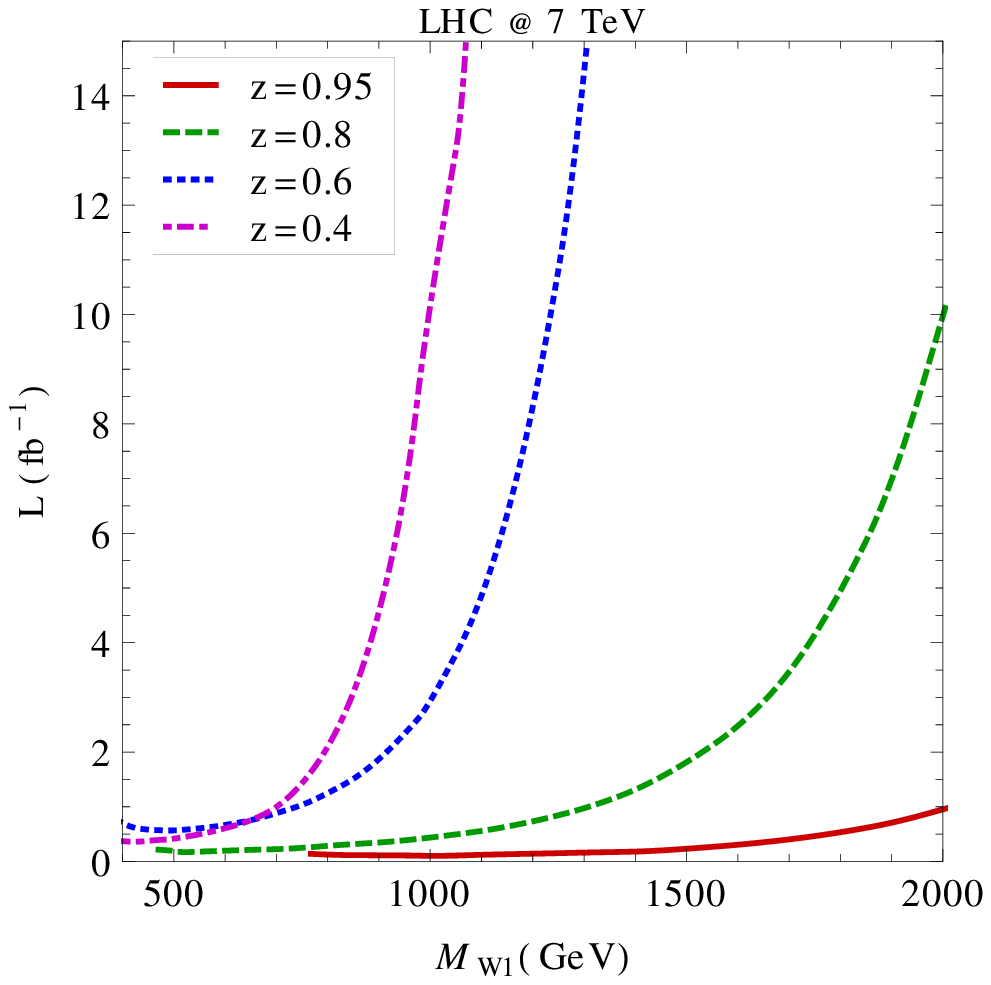,width=7.5cm}}
\put(3.5,2.7){\epsfig{file=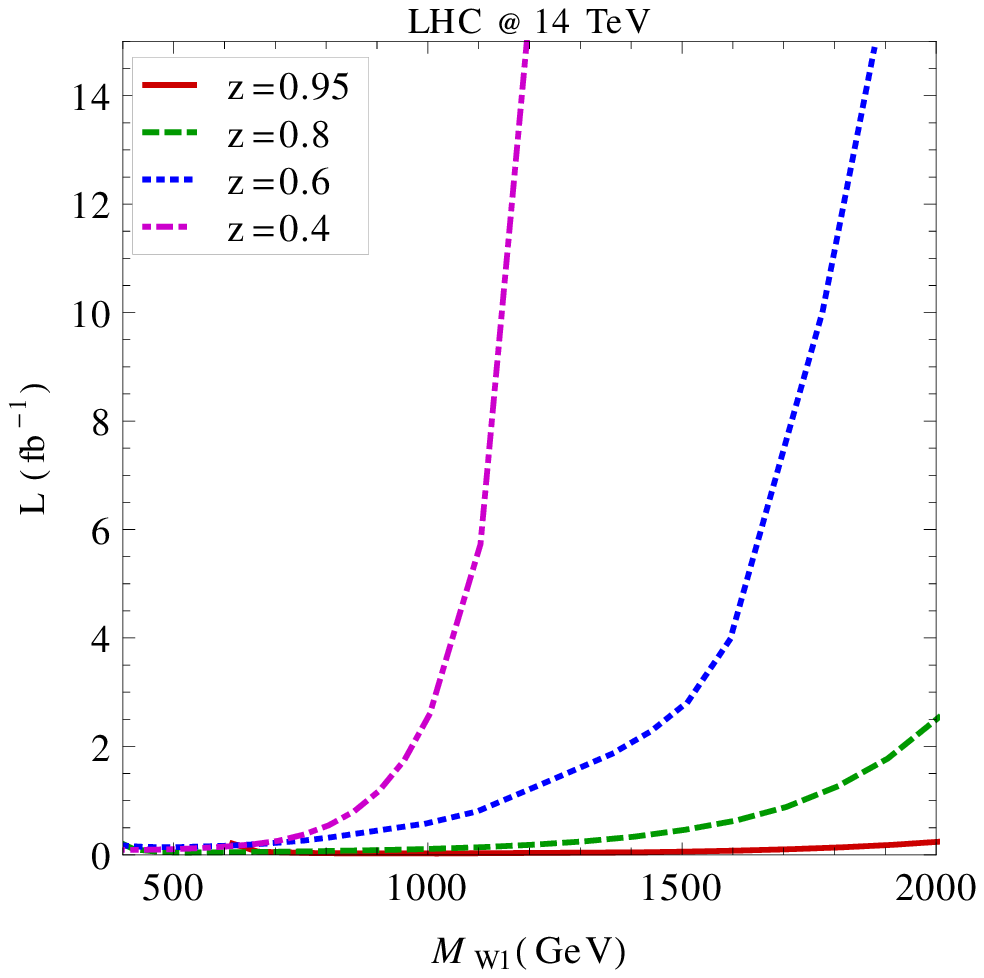,width=7.5cm}}
\end{picture}
\end{center}
\vskip -3.5cm
\caption{Left: Minimum luminosity needed for a 5$\sigma$-discovery of $W_{1,2}$-bosons
at the 7 TeV LHC, assuming maximal $W_{1,2}$-boson couplings to 
SM fermions and $z$=0.4, 0.6, 0.8, 0.95. Right: same curves at the upgraded 14 TeV LHC.}
\label{fig:min_luminosity}
\end{figure}

To conclude, we show in Fig. \ref{fig:min_luminosity} the minimum luminosity 
needed to claim a $W_{1,2}$-boson discovery during the early stage 
(left panel) and project stage (right panel) of the LHC.

\section{Conclusions}
\label{conclusions}

In this paper, we have studied the phenomenology of the next-to-minimal 
(or 4-site) Higgsless model, which extends the minimal (or 3-site) 
Higgsless model by including one additional site in the linear moose picture. 
The 4-site model is a deconstructed theory based on the 
$SU(2)_L\times SU(2)_1\times SU(2)_2\times U(1)_Y$ gauge symmetry. It predicts 
four charged and two neutral extra gauge bosons: $W^\pm_{1,2}$ and $Z_{1,2}$. 
We have focused on the properties of the charged gauge sector and on its 
discovery prospects at the 7 TeV LHC in the final state with one isolated 
electron (or positron) and large missing energy during the next year.

The phenomenology of the 4-site model is controlled by three free parameters
beyond the SM ones: the two extra gauge boson masses, $M_{W1,W2}$ and the 
coupling between the lighter extra gauge boson and SM fermions, $a_{W1}$. At 
fixed ratio $z=M_1/M_2\simeq M_{W1}/M_{W2}$, the model can be thus visualized 
in the ($M_{W1}, a_{W1}$) plane.  
The mass spectrum is bounded from above by the requirement of perturbative 
unitarity, and from below by EWPT. The viable mass range is thus around  
[250, 3000] GeV. EWPT also constrain the magnitude of the couplings. 
Nevertheless, the allowed parameter space is sizeable. This represents the 
major 
novelty of the 4-site model which, in contrast to its minimal version (or 
three site model), can solve the dichotomy between unitarity and EWPT bounds 
without imposing the extra vector bosons to be fermiophobic. As a consequence, 
the Drell-Yan process becomes a relevant channel for the direct search of 
extra gauge bosons at the LHC.

In this paper, we have first described the main properties of the new 
$W_{1,2}$-bosons. We have shown their total decay widths and branching ratios 
into fermions and bosons. The results can be summarized in two main points. 
First, BRs into lepton pairs and boson pairs decaying in turn into leptons 
can compete when looking at purely leptonic final states. Within the 
4-site Higgsless model, the DY process is thus a strong discovery channel. 
Second, in contrast to common $W^\prime$ bosons which purely decay into SM 
fermions and appear as narrow resonances, the $W_{1,2}$ bosons can display a 
broad peaking structure. This feature poses one more challenge to the  
experimental analyses commonly based on the assumption of narrow resonances. 

In order to show what signature might be expected, we have presented four 
sample scenarios corresponding to different points in the parameter space. 
In the 4-site model, no relation between lighter and heavier extra gauge 
bosons is predicted ($z=M_1/M_2$ is a free parameter). So, for sake of 
completeness, we have chosen four cases where in principle the two 
$W_{1,2}$-bosons could appear as resonances either quite distant in mass or
almost degenerate. We have then computed the limits on $W_{1,2}$-boson 
masses and couplings from direct searches in both neutral and charged 
Drell-Yan channels at the Tevatron with integrated luminosity L=10 fb$^{-1}$. 
The outcome is that the 4-site model is weakly bounded. For low $z$ values, 
only a small corner around the minimum mass, $M_{W1}\ge 350$ GeV, is excluded.
For high $z$ values, one can exclude the model up to $M_{W1}\simeq 1100$ GeV   
assuming maximal couplings between extra gauge bosons and SM fermions.
The present data collected at the LHC, which has now over 1 fb$^{-1}$, 
extend the Tevatron limits by several hundreds of GeV. In the low edge of the 
$z$ interval, the exclusion limit can reach $M_{W1}\simeq 800$ GeV. It further 
grows with increasing $z$, and can even exclude the entire mass spectrum in 
the almost degenerate scenario ($z$=0.95), not only for maximal couplings.

Looking at one-year timescale, we have shown how the expected 7 TeV LHC with 
L=10 fb$^{−1}$ could sensibly extend the 4-site physics search to regions 
of the parameter space with smaller $W_{1,2}$-boson couplings to ordinary 
matter. The presented results are of course preliminary, coming from a pure 
parton level analysis. They nevertheless show that the 4-site model can be
tested already during the early LHC stage in a large portion of its parameter
space.

\noindent
{\bf Note added in proof.}
Soon after the publication of our paper, the new results on the search for a 
new heavy gauge boson $W^\prime$ decaying to a charged lepton (muon or 
electron) and a neutrino at the LHC were published by the CMS experiment 
\cite{cms:last}. The experimental analysis is based on the data collected in 
2011, which 
correspond to an integrated luminosity of 1.1 fb-1. As both setup and signal 
definition are different, we cannot directly compare the results of our paper 
with the limit placed by CMS. \\

\begin{acknowledgments}
E.A. and D.B. acknowledge financial support from the NExT Institute and 
SEPnet. E.A. thanks the theoretical physics department of the University of 
Torino for hospitality. 
The work of S.D.C., D.D. and L.F. is partly supported by the Italian Ministero
dell'Istruzione, dell' Universit\`a
e della Ricerca Scientifica, under the COFIN program (PRIN 2008).\\
\end{acknowledgments}

\appendix

%\addcontentsline{toc}{chapter}{References}
%\bibliographystyle{unsrt}
\bibliography{bib_last_2}

\end{document}